# Dynamic Binary Translation for SGX Enclaves

Jinhua Cui
National University of Defense Technology
National University of Singapore
jhcui.gid@gmail.com

Shweta Shinde*
ETH Zurich
shweta.shivajishinde@inf.ethz.ch

Satyaki Sen
National University of Singapore
satyakisen2012@gmail.com

Prateek Saxena
National University of Singapore
prateeks@comp.nus.edu.sg

Pinghai Yuan
National University of Singapore
pinghaiyuan@gmail.com

## ABSTRACT

Enclaves, such as those enabled by Intel SGX, offer a hardware primitive for shielding user-level applications from the OS. While enclaves are a useful starting point, code running in the enclave requires additional checks whenever control or data is transferred to/from the untrusted OS. The enclave-OS interface on SGX, however, can be extremely large if we wish to run existing unmodified binaries inside enclaves. This paper presents RATEL, a dynamic binary translation engine running inside SGX enclaves on Linux. RATEL offers *complete interposition*, the ability to interpose on all executed instructions in the enclave and monitor all interactions with the OS. Instruction-level interposition offers a general foundation for implementing a large variety of inline security monitors in the future.

We take a principled approach in explaining why complete interposition on SGX is challenging. We draw attention to 5 design decisions in SGX that create fundamental trade-offs between performance and ensuring complete interposition, and we explain how to resolve them in the favor of complete interposition. To illustrate the utility of the RATEL framework, we present the first attempt to offer *binary compatibility* with existing software on SGX. We report that RATEL offers binary compatibility with over 200 programs we tested, including micro-benchmarks and real applications such as Linux shell utilities. Runtimes for two programming languages, namely Python and R, tested with standard benchmarks work out-of-the-box on RATEL without any specialized handling.

## 1 INTRODUCTION

Commercial processors today have native support for trusted execution environments (TEEs) to run user-level applications in isolation from other software on the system. A prime example of such a TEE is Intel Software Guard eXtensions (SGX) [56]. The hardware-isolated environment created by SGX, commonly referred to as an *enclave*, runs a user-level application without trusting privileged software. Enclaves offer a good basis for isolation, as they do not necessarily place trust on the OS and allow us to restrict the code base to trust. Further, they open up the possibility of reverse sandboxing, where the enclaved application protects itself from attacks arising from the OS [46].

SGX exposes an extremely large interface between the enclave and the OS, including the potential to transfer control to the OS at every memory access (e.g. via memory faults) or instruction executed (e.g. via timer interrupts and exceptions). Furthermore, the demand for running commodity applications inside SGX has surged, but these applications are not written to deal with the threat of a malicious OS. Therefore, the ability to interpose on all control and data passed on the enclave-OS interface is an important building block. Such interposition can be used for implementing compatibility frameworks, a host of well-known inline security monitors, and sandboxing techniques inside enclaves [13, 44, 70, 81, 82].

Complete interposition on enclave-OS interface is a known challenge. For example, a long line of work on frameworks which aim to run existing software on SGX highlights the difficulty of ensuring compatibility [15, 19, 28, 69, 72]. In this work, we address this challenge by taking a new approach: we enable *dynamic binary translation* (DBT), i.e., the ability to interpose on all enclave instructions executed in the enclave. Our work enables DBT on Intel SGX enclaves for unmodified x86_64 Linux binaries by designing a system called RATEL. RATEL is available open-source [1] and it builds on DynamoRIO, an industrial-strength DBT engine originally designed for non-enclave code [23]. The RATEL DBT engine does *not* trust the OS, and enclave applications running on RATEL are assumed to be unaware of its presence. A security monitor implemented using RATEL can mediate and intercept on all instructions, entry-exits, system calls, dynamically generated code, asynchronous events, virtual address accesses, and run-time loading of code and data in the enclave—a foundation for implementing a wide variety of security-related instrumentation on enclaves in the future, without specializing to individual applications or language runtimes.

To illustrate one advantage of such seamless interposition, in this work, we use RATEL to build a binary compatibility layer for SGX. Binary compatibility creates the illusion for an unmodified application binary as if it is running in a normal OS process, rather than in a restricted environment such as an enclave. In designing this layer, we observe several trade-offs arise between ensuring *complete interposition* on the OS-enclave boundary and the resulting performance. These trade-offs are orthogonal to security concerns pointed out in prior works (c.f. Iago attacks [29], side-channels [88]). We observe that these trade-offs are somewhat fundamental and rooted in 5 specific restrictions imposed by the SGX design, which create sweeping incompatibility with multi-threading, memory mapping, synchronization, signal-handling, shared memory, and other commodity OS abstractions. Our design resolves these trade-offs consistently in the favor of complete interposition rather than performance. In this sense, our work departs from prior works.

RATEL is the first system that enables DBT and binary compatibility for SGX, to the best of our knowledge. Prior works have proposed a number of different ways of achieving partial compatibility—offering specific programming languages for authoring enclave code [16, 61, 84], keeping compatibility with container interfaces [15, 46], or conformance to specific versions of library interfaces provided by library OSes [19, 28, 65, 69, 72]. All of these designs, however, assume that the application binaries run benign code that uses a particular prescribed interface to achieve compatibility on SGX—for example, application binaries are expected to be relinked against specific versions of libraries (e.g., `musl`, `libc`, `glibc`), ported to a customized OS, or containerized. In contrast, RATEL interposes at the instruction-level execution of unmodified program binaries in the enclave, and this approach conceptually does not require such strong assumptions.

**Results.** We highlight 3 results showing the egalitarian compatibility offered by RATEL. First, we find that RATEL supports more than one language runtimes (e.g. Python and R) out-of-the-box, without requiring any language-specific design decisions. Second, we successfully run a total of 203 unique unmodified binaries across 5 benchmark suites (58 binaries), 4 real-world application use-cases (12 binaries), and 133 Linux utilities. These encompass various workload profiles including CPU-intensive (SPEC 2006), I/O system call intensive (FSCQ, IOZone), system stress-testing (HBenchOS), multi-threading support (Parsec-SPLASH2), a machine learning library (Torch), and real-world applications demonstrated in prior works on SGX. RATEL offers compatibility but does not force applications to use any specific libraries or higher-level interfaces. At the same time, our presented techniques work without any specialization per target application or runtime, highlighting that DBT can be a general solution to compatibility on enclaves. Lastly, we show that RATEL has comparable[1] or better compatibility than Graphene-SGX, which requires relinking with a particular version of `libc`, and is one of the longest maintained SGX compatibility infrastructure available publicly.

## 2  WHY IS COMPLETE INTERPOSITION CHALLENGING?

Intel SGX allows execution of code inside a hardware-isolated environment called an enclave for running user-level application code [33].[2] Our goal is to interpose on all the instructions executed inside the enclave. This is a challenge on SGX because of its severe threat model and restrictions placed on enclave code. The OS is not trusted in this threat model. SGX enforces confidentiality and integrity of enclave-bound code and data. All enclave memory is private and only accessible when executing in enclave-mode. Data exchanged with the external world (e.g., the host application or OS) must reside in public memory which is not protected. At runtime, execution control can only synchronously enter an enclave via `ECALLs` and exit an enclave via `OCALLs`, which are primary interfaces provided by SGX for effecting system calls (syscalls). Any illegal instructions or exceptions in the enclave create asynchronous entry-exit points. SGX restricts these to pre-specified points in the program. If the enclave execution is interrupted asynchronously, SGX saves the enclave code execution context and resumes it at the entry point later [2].

---

[1] We chose not to support fork in RATEL at the moment [18].

[2] Unless stated otherwise, we use the term Intel SGX v1 to refer to the hardware as well as the trusted platform software (PSW) and the trusted software development kit (SDK), as shown in Figure 2.

| OS abstraction | Restrictions affecting abstraction |
|---|---|
| System call arguments | R1 |
| Dynamic loaded/generated code | R2 |
| Thread support | R5, R2 |
| Signal handling | R1, R5 |
| Thread synchronization | R3, R1 |
| File/memory mapping | R1, R2, R3, R4 |
| IPC/shared memory | R3, R4 |

**Table 1: Ramifications of SGX design restrictions on common OS abstractions.**

### 2.1  Restrictions Imposed by SGX Design

Intel SGX protects the enclave by enforcing strict isolation at several points of interactions between the OS and the user enclave code. We outline 5 restrictions that the design of SGX imposes:

**R1. Spatial memory partitioning.** SGX enforces *spatial* memory partitioning. It reserves a region that is private to the enclave and the rest of the virtual memory is public. Memory can either be public or private, not both.

**R2. Static memory partitioning.** The enclave has to specify the spatial partitioning *statically*. The size, type (e.g., code, data, stack, heap), and permissions for its private memory have to be specified before creation and these attributes cannot be changed at runtime.

**R3. Non-shareable private memory.** An enclave cannot share its private memory with other enclaves on the same machine.

**R4. 1-to-1 private virtual memory mappings.** Private memory spans over a contiguous virtual address (VA) range, the start address of which is decided by the OS. The private VA space has a 1-to-1 mapping with the physical address (PA) space.

**R5. Fixed entry points.** Enclave can resume execution only from its last point and context of exit. Any other entry points/contexts have to be statically pre-specified in the binary as valid ahead of time.

### 2.2  Ramifications on Incompatibility

Restrictions R1-R5 are a systematic way to understand the incompatibility created by design choices in SGX with the OS and application functionality. Table 1 summarizes the effect.

**R1.** Since SGX spatially partitions the enclave memory, any data which is exchanged with the OS requires copying between private and public memory. In normal applications, an OS assumes that it can access all the memory of a user process, but this is no longer true for enclaves. Any syscall arguments that reside in enclave private memory are not accessible to the OS or the host process. The enclave has to explicitly manage a public and a private copy of the data to make it accessible externally and to shield it from unwanted modification when necessary. We refer to this as a *two-copy mechanism*. Thus, *R1* breaks functionality (e.g., system calls, signal handling, futex), introduces non-transparency (e.g., explicitly synchronizing both copies), and introduces security gaps (e.g., TOCTOU attacks [29, 40]).

**R2.** Applications often require changes to the size or permissions of enclave memory. For example, memory permissions change after dynamic loading of libraries (e.g., dlopen) or files (e.g., mmap), executing dynamically generated code, creating read-only zeroed data segments (e.g., .bss), and for software-based isolation of security-sensitive data. The restriction R2 is incompatible with such functionality. To work with this restriction, applications require careful semantic changes: either weaken the protection (e.g., read-and-execute instead of read-or-execute), use the two-copy mechanism, or rely on some additional form of isolation (e.g., using segmentation or software instrumentation).

**R3.** SGX has no mechanism to allow two enclaves to share parts of their private memory directly. This restriction is incompatible with the synchronization primitives like locks and shared memory when there is no trusted OS synchronization service. Keeping 2 copies of a shared lock breaks its semantics and creates a chicken-and-egg issue: how to synchronize the 2 copies without another trusted synchronization primitive.

**R4.** When applications demand new virtual address mappings (e.g., malloc), the OS adds these mappings. Normally, applications can ask the OS to map the same physical page at several different offsets, either with same or different permissions—for example, say when the same file is mapped as read-only at two places in the program space. On SGX, however, the same PA cannot be mapped to multiple enclave VAs. Any such mappings lead to memory protection faults.

**R5.** SGX starts or resumes enclave execution only from controlled entry points, i.e., which have to be statically identified virtual addresses. However, there are several unexpected entry points to an application when we run them unmodified in an enclave (e.g., exception handlers, library functions, illegal instructions). Statically determining all potential program points for re-entry is difficult. Moreover, when the control enters back into the enclave after an exit, SGX requires that the program execution context at the time of re-entry and exit should be the same. This does not adhere to typical program functionality. Normally, if the program wants to execute custom error handling code, say after a divide-by-zero (SIGFPE) or illegal instruction (SIGILL), it can resume execution at a handler function in the binary with appropriate execution context setup by the OS. On the contrary, SGX will resume enclave execution at the same instruction and same context (not the OS setup context for exception handling), thus re-triggering the exception.

The above restrictions are specific to Intel SGX v1. Intel has proposed SGX v2 wherein an enclave can make dynamic changes to private page permissions, type, and size. Note that SGX v2 only addresses R2 partially, while all the other restrictions still apply to SGX v2. Thus, for the rest of the paper, we describe our design based on SGX v1. We discuss their specific differences to v2 and its ramification in Section 7.

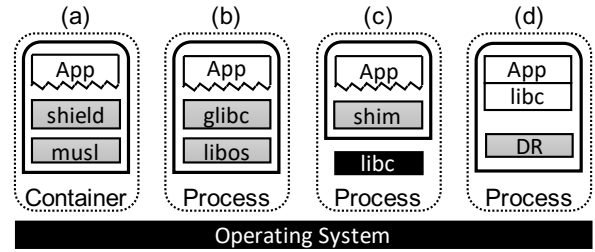

**Figure 1: Different abstraction layer choices for compatibility. Black shaded regions are untrusted, gray shaded regions are modifications or additions, thick solid lines are enclave boundaries, dotted lines are container/process boundaries, clear boxes are unmodified components, zig-zag lines show break in compatibility. (a) Container abstraction with musl or libc interface (Scone [15]). (b) Library OS with glibc interface (Graphene-SGX [28]). (c) Process abstraction with POSIX interface (Panoply [72]). (d) Dynamic Binary Translation with DynamoRIO in RATEL (This work).**

## 3 OVERVIEW

Our work poses the following question: Can complete interposition on the OS-enclave interfaces be achieved on the SGX platform? We present the first system that allows interposing on all enclave-bound instructions, by enabling a widely-used dynamic binary translation (DBT) engine inside SGX enclaves. Our system is called RATEL.

Before we present the design of our DBT engine, we emphasize a key design trade-off: *Working with restrictions R1 − R5, we observe that one is forced to choose between complete interposition at the OS-enclave interfaces and performance.* We explain these trade-offs in Section 3.2. Our design picks completeness of interposition over performance, wherever necessary. In this design principle, it fundamentally departs from prior work.

Several different approaches to enable applications in SGX enclaves have been proposed. In nearly all prior works, performance consideration dominates design decisions. A prominent way to side-step the performance costs of ensuring compatibility is to ask the application to use a prescribed program-level interface or API. The choice of interfaces varies. They include specific programming languages [31, 38, 41, 84], application frameworks [50], container interfaces [15], and particular implementation of standard libc interfaces. Figure 1 shows the prescribed interfaces in three approaches, including library OSes and container engines, and where they intercept the application to maintain compatibility. Given that complete instruction-level interposition is not the objective of prior works, they handle only subsets of *R1 − R5*. One drawback of these approaches is that if an application does not originally use the prescribed API, the application needs to be rewritten, recompiled from source, or relinked against specific libraries. Further, enclave programs may invoke the OS interfaces directly outside the prescribed API. The approach of complete interposition at the lowest level of interfaces (i.e. at each executed instruction) offers a powerful way of providing compatibility *without* specializing to specific target applications or making any such assumptions on the application behavior.

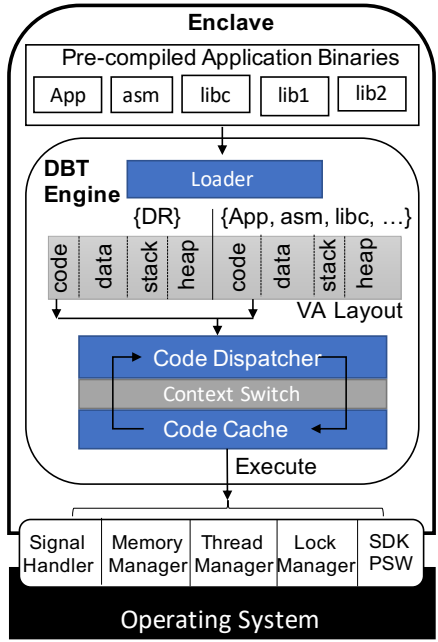

**Figure 2:** RATEL **overview.**

As an illustration of its utility, we show that we can build a binary compatibility layer for Linux-based SGX enclaves on RATEL. Application binaries are originally created with the intention of running on a particular OS in an unrestricted OS process environment. A binary compatibility layer runs below the application and translates any code illegal in the restrictive SGX environment to the appropriate enclave-OS interfaces. In concept, application code is thus free to use any library, direct assembly code, and runtime that uses the Linux system call interfaces. Furthermore, RATEL has about 26 additional instruction-level runtime profilers and monitors, which are pre-existing in our baseline DBT engine (see Section 6.4). These become available to applications running on SGX enclaves directly. Such instrumentation can be used for debugging, resource accounting, or implementing inline security monitors for enclaved code in the future.

### 3.1 Background on DBT

Dynamic binary translation (DBT) is a well-known approach to binary code instrumentation and implementing inline reference monitors. It intercepts each instruction in a program before it executes [49]. In this paper, we choose DynamoRIO as our DBT engine, since it is open-source and widely used in industry [23].[3] Vanilla

---
[3] Another option is Intel Pin [47], but it is not open-source.

DynamoRIO works much like a just-in-time compilation engine which dynamically re-generates (and instruments) code of the application running on it. At a high level, DynamoRIO first loads itself and then loads the application code in a separate part of the VA space, as shown in Figure 2. Similarly, it sets up two different contexts, one for itself and one for the application. DynamoRIO can update the code on-the-fly before putting it in the code cache by re-writing instructions (e.g., convert a syscall instruction to a stub or library function call). Such rewriting ensures that DynamoRIO engine takes control before each block of code executes, enabling the ability to interpose on every instruction. Instrumented code blocks are placed in a region of memory called a code cache. When the code cache executes, DynamoRIO regains control as the instrumentation logic desires. It does post-execution updates to itself for book-keeping or to the program's state. Additionally, DynamoRIO hooks on all events intended for the process (e.g., signals). The application itself is prevented from accessing DynamoRIO memory via address-space isolation techniques [49]. Thus, it acts as an arbiter between the application's binary code and the external environment (e.g., OS, filesystem) with complete interposition.

The original DynamoRIO engine is designed to work for non-enclave code. We adapt it to work inside SGX enclaves, resulting in our RATEL system. To contrast it with the approach of changing `libc`, DBT intercepts the application right at the point at which it interacts with the OS (Figure 1) for SGX compatibility. RATEL retains the entire low-level instruction translation and introspection machinery of DynamoRIO, including the code cache and its performance optimizations. This enables reusing well-established techniques for application instrumentation and performance enhancements. We eliminate the support for auxiliary client plugins to reduce TCB, but a suite of built-in runtime profilers (see Table 10) which do not use the DynamoRIO client plugin interfaces are retained in RATEL.

### 3.2 RATEL Approach

RATEL provides compatibility for both the DynamoRIO DBT engine as well as any application binary code that runs translated. We provide a high-level overview of our design and explain the key trade-offs this design makes.

**High-level Overview.** As a first step, RATEL loads the DynamoRIO dynamic translation engine at a specific location in virtual memory, which we denote as $A$. Let us say that the vanilla dynamic translation engine in DynamoRIO is coded to access virtual address memory regions denoted as $B$ and the target application accesses regions $C$ respectively. The basic principle behind the design of RATEL is to ensure *referential transparency*: whenever the DynamoRIO engine accesses any virtual address that would have been a location in $B$ (without RATEL), it must now access the corresponding location in $A$ in RATEL. Similarly, if the application would have accessed a location in $C$ originally, RATEL must ensure that it accesses the corresponding to its translated location. To do this, RATEL must (a) intercept all operations that create virtual memory maps (e.g., via static and dynamic loading), and (b) keeps an address translation table in $A$ for translating program accesses made by the target application dynamically. Note that such referential transparency for memory accesses provides compatibility with position-independent code, dynamically generated code, and shadow memory data structures (e.g., shadow stacks) that the application may have originally used—the memory references at runtime resolve consistently to the same translated address and the values read/written as thus consistent with the original run. For security, Ratel must ensure that all accesses to $A$ originate from the dynamic translation engine itself, and the application code is unable to access $A$ directly—a memory isolation policy. Ratel enforces this policy for the dynamic translation engine by modifying its code statically. For the target application, memory isolation can be enforced at runtime through the instruction rewriting capability of DynamoRIO itself, as done in program shepherding [49].

In addition, Ratel modifies DynamoRIO to adhere to SGX virtual memory limitations (R1-R4). In designing Ratel, we statically change the DynamoRIO code to load it at a fixed memory region $A$. Note that $A$ can be fixed at the time of initialization of the process (loading), therefore, it does *not* break compatibility with address-space layout randomization. This allows us to load Ratel and start its execution without violating the memory semantics of SGX. We register a fixed entry point in Ratel when entering or resuming the enclave. This entry point acts as a unified trampoline, such that upon entry, Ratel decides where to redirect the control flow, depending on the previously saved context. In DynamoRIO code, we statically replace all instructions that are illegal in SGX with an external call that executes outside the enclave. Thus, Ratel execution itself is guaranteed to never violate R5.

Ratel has complete control over the loading and running of the translated application binary. Therefore, to ensure that the application adheres to R1-R5, Ratel dynamically rewrites the instructions before they are executed from the code cache. To keep compatibility with R2, we statically initialize the virtual memory size of the application to the maximum allowed by SGX; the type and permissions of memory are set to the specified type in the original binary. Ratel augments its memory manager to keep track of and transparently update the application memory layout as it changes during execution. At runtime, the application can make direct changes to its own virtual memory layout via system calls. Ratel dynamically adapts these changes to SGX by making two copies, wherever necessary, or by relocating the virtual address regions. Ratel intercepts all application interactions with the OS. It modifies application parameters, OS return values and events for monitoring indirect changes to the memory (e.g., thread creation). Before executing any application logic, Ratel scans the code cache for any instructions (e.g., `syscall`, `cpuid`) that may potentially be deemed as illegal in SGX and replaces it with an external call. In the other direction, Ratel also intercepts OS events on the behalf of the application. Upon re-entry, if the event has to be delivered to the application (e.g., signals for application itself), it sets/restores the appropriate execution context and resumes execution via the trampoline. In this way, Ratel remedies the application on-the-fly to adhere to R1-R5.

**Resolving Key Design Trade-offs.** Ratel helps to interpose on the enclave code without relying on the untrusted OS. In doing so, the SGX restrictions $R1 - R5$ give rise to trade-offs between ensuring complete interposition and having low performance overheads. We point out that these are somewhat fundamental and apply to Ratel and other compatibility efforts equally. However, Ratel chooses completeness in its interposition over performance, whenever conflicts arise.

Due to $R1$, whenever the application wants to read from or write data outside the enclave, the data needs to be placed in public memory. Computing on data in public memory, which is exposed to the OS, is insecure. Therefore, if the application wishes to securely compute on the data, a copy must necessarily be maintained in a separate private memory space, as $R2$ forbids making changes to the memory permissions dynamically. This leads to a "two copy" mechanism, instances of which repeat throughout the design. The two-copy mechanism, however, incurs both space and computational performance overheads, as data has to be relocated at runtime.

$R3$ creates an "all or none" trust model for enclaves. Either memory is shared with all entities (including the OS) or kept private to one enclave. $R4$ restricts sharing memory within an enclave further. These restrictions conflict with semantics of shared memory and synchronization primitives. For instance, synchronization primitives such as `futexes` are implemented with a single memory copy that the OS is trusted to manage securely—such a design is in direct conflict with the SGX security model. To implement such abstractions securely, designs on SGX must rely on a trusted software manager which necessarily resides in an enclave, since the OS is untrusted (see Section 4.4). Applications can then regain compatibility with locks and shared memory abstractions, but at the cost of performance: Access to shared memory or synchronization primitives turn into (possibly remote) procedure calls to the trusted manager enclave.

Restriction $R5$ requires that whenever the enclave resumes control after an exit, the enclave state (or context) should be the same as right before exit. This implies that the security monitor (e.g., the DBT engine) must take control before all exit points and after resumption, to save-restore contexts—otherwise, the interposition can be incomplete, creating security holes and incompatibility. Without guarantees of complete interposition, the OS can return control into the enclave, bypassing security checks that the DBT engine implements. The price for complete interposition on binaries is performance—the DBT engine must intercept all entry/exit points and simulate additional context switches in software. Prior approaches, such as library OSes, choose performance over completeness in interposition, by asking applications to link against specific library interfaces which constrict enclave-OS interaction via certain specified library interfaces. But, this does not enforce complete interposition. Applications, due to bugs or when exploited, can make direct OS interactions without using the prescribed API, use inline assembly, or override entry handlers setup by the library OS. All scenarios in which applications go outside the prescribed interfaces, the library OS design requires special handling.

Several additional security considerations arise in the implementation details of our design. These include (a) avoiding naïve designs that have TOCTOU attacks; (b) saving and restoring the execution context from private memory; (c) maintaining Ratel-specific metadata in private memory to ensure integrity of memory mappings that change at runtime; and (d) explicitly zeroing out memory content and pointers after use. We explain them inline in Section 4.

## 3.3 Threat Model & Scope

RATEL is best viewed as a general framework for instruction-level interposition on enclaved binaries, rather than a stand-alone sandboxing engine that protects enclaves against all possible OS attacks. RATEL itself does *not* trust the OS or any security guarantees the OS provides. Binaries running on top of RATEL otherwise follow the same threat model as vanilla DBT engines [80]. Application binaries are assumed to not be aware of the presence of RATEL—they are benign binaries but which can be exploited via externally-provided inputs. Under exploitation, malicious code may execute on RATEL. RATEL provides instruction-level instrumentation of all instructions executed and does not provide any higher-level security guarantees beyond that. For example, malicious code can readily determine that it is running on RATEL [39] and, therefore, RATEL is not suitable for analyzing analysis-evading malicious code. Using the instruction-level monitoring capability, the vanilla DynamoRIO engine itself provides certain built-in security mechanisms, which are preserved in RATEL. Specifically, ASLR for all heap regions is turned on and the code cache is randomized. RATEL isolates the stack, dynamically allocated memory, and file-mapped I/O regions through instrumentation. The main new challenges highlighted in this work are those due to enabling DBT on SGX, while most other threats to DBT-based instrumentation are pre-existing and known. The design trade-offs we emphasize apply to any interposition framework that runs on SGX—but these have remained largely implicit (not unstated) in prior compatibility works.

Building an end-to-end secure sandbox on top of RATEL requires additional security mechanisms, which are common to other systems and are previously known. These mechanisms include encryption/decryption of external file or I/O content [15, 46, 69, 77], sanitization of OS inputs to prevent Iago attacks [29, 48, 74, 79], defenses against known side-channel attacks [21, 45, 63, 70, 71], additional attestation or integrity of dynamically loaded/generated code [41–43, 83], and so on. These are important but largely orthogonal to our focus.

Our binary compatibility layer has support for large majority but not all of the Linux system calls. The most notable of these unsupported system calls is fork which is used for multi-processing. Note that the basic design of RATEL can be extended to support fork with the two-copy mechanism, similar to prior work [72]. However, maintaining compatibility with fork blindly is a questionable design decision, especially for enclaves, as has been argued extensively [18]. A recent work on SGX compatibility has left out support for fork based on the same observation [69].

## 4 RATEL DESIGN

We explain how RATEL handles syscalls, memory, threads, synchronization, and exceptions/signals inside SGX enclaves in the presence of restrictions $R1 - R5$.

## 4.1 Syscalls & Unanticipated Entry-Exits

SGX does not allow enclaves to execute several instructions such as syscall, cpuid, and rdtsc. If the enclave executes them, SGX exits the enclave and generates a SIGILL signal. Gracefully recovering from the failure requires re-entering the enclave at a different program point. Due to R5, this is disallowed by SGX. In RATEL, either DynamoRIO or the application can invoke illegal instructions, which may create unanticipated exits from the enclave.

RATEL changes DynamoRIO logic to convert such illegal instruction to stubs that either delegate or emulate the functionality. For the target application, whenever RATEL observes an illegal instruction in the code cache, it replaces the instruction with a call to the RATEL syscall handler function. RATEL has three ways of handling system call execution:

(1) *Complete delegation*: Entirely delegate the syscall instruction and handler to code outside the enclave;

(2) *Partial delegation*: Execute the syscall instruction outside, and then update the private in-enclave state; or

(3) *Emulation*: Completely simulating the syscall behavior with a handler inside the enclave.

RATEL uses complete delegation for file, networking, and timer related system calls. It uses partial delegation for memory management, threads, and signal handling. We outline the details of other syscall subsystems that are fully or partially emulated by RATEL in Sections 4.2, 4.3, 4.4, and 4.5. RATEL uses emulation for very few system calls. For example, the arch_prctl syscall is used to read the FS base. RATEL emulates it by executing a rdfsbase instruction.

**Creating and Synchronizing Memory Copies.** Syscalls access process memory for input-output parameters and error codes. Since enclaves do not allow this, for delegating the syscall outside the enclave, RATEL creates a copy of input parameters from private memory to public memory. This includes simple value copies as well as deep copies of structures. The OS then executes the syscall and generates results in public memory. After the syscall completes, RATEL copies back the OS-provided return values and error codes to private memory.

Memory copies alone are not sufficient. For example, when loading a library, the application uses dl_open which in turn calls mmap, which must execute outside the enclave. Thus the mmap call outside the enclave will map the library in the untrusted public address space of the application. However, the original intent of the application is to map the library inside the enclave private memory. As another example, consider when the enclave code wants to create a new thread local storage (TLS) segment. Due to the restrictive SGX environment, RATEL must execute the system call outside the enclave, and the new thread is created for the DynamoRIO runtime instead of the target application. In all such cases, RATEL takes care to explicitly propagate changes to inside the enclave, i.e. reflect changes to private memory.

**Checking Memory State after Syscalls.** RATEL resumes execution in the enclave only after the syscall state has been completely copied inside the enclave. This allows it to employ sanitization of OS return values before using it. Previously known sanitization checks for Iago attacks can be implemented here [74]. Note that all such sanitization checks must execute inside the enclave and *after* the state from the public memory is copied into the enclave private memory. This caveat is important to avoid TOCTOU attacks wherein the OS modifies public memory state before or midway through the execution of the sanitization checks.

## 4.2 Memory Management

Ratel utilizes partial emulation for syscalls that change the process virtual memory layouts and permissions (e.g., `mmap`, `mprotect`, `fsync`, and so on). It executes the syscall outside the enclave and then explicitly reflects the memory layout changes inside the enclave. First, this is not straightforward. Due to R1-R4, several layout configurations are not allowed for enclave virtual memory (e.g., changing memory permissions). Second, Ratel does not trust the OS information (e.g., via `procmap`). Hence, Ratel must use a two-copy mechanism when it uses the partial delegation approach.

Specifically, Ratel maintains its own `procmap`-like structure to keep its own view of the process virtual memory inside the enclave, tracks the memory-related events, and updates the enclave state. For example in the case of `mmap` syscall to map a file in enclave private memory, the handler outside the enclave creates a public memory region by invoking the OS itself. Then, Ratel allocates a region of private memory which mirrors the content of the file mapped outside the enclave, and updates its internal `procmap`-like structure to record the new virtual addresses created. Further, Ratel synchronizes the two-copies of memories to maintain execution semantics on all subsequent changes to mmapped-memory. This is done whenever the application unmaps the memory or invokes the `sync`/`fsync` syscalls.

Ratel does not blindly replicate OS-dictated memory layout changes inside the enclave. It first checks if the resultant layout will violate any security semantics (e.g., mapping a buffer to zero-address). It proceeds to update enclave layout and memory content only if these checks succeed. To do this, Ratel keeps its metadata in private memory.

With interposition over memory management, Ratel transparently side-steps SGX restriction due to R2. When application makes changes to the permissions of a memory region (say *X*) dynamically, Ratel moves the content to a memory region (say *Y*) which has the required permissions. To do this, Ratel requires a "stash" of unused private memory regions (*Y*) that are originally not used by the application. These memory regions are allocated statically by Ratel at the start and their page permissions are set to readable, writable, and executable. Subsequently, when the application binary accesses memory *X*, Ratel dynamically translates the access to the copy in memory *Y*. This allows Ratel to transparently emulate the permission change, which is otherwise a disallowed behavior inside the enclave.

Note that the stash pool of private memory regions *Y* are private to the enclave, but have overly permissive access rights. This can be avoided by reserving regions with write-only and execute-only permissions, but this may decrease the size of usable non-stash memory. Alternatively, the access rights can be implemented through runtime monitoring of memory accesses, but this adds performance costs. The trade-offs are inherent to restrictions *R*1 and *R*2.

## 4.3 Multi-threading

As per restriction *R*2, SGX requires the application to pre-declare the maximum number of threads before execution. Further, it does not allow the enclave to resume at arbitrary program points or execution contexts, as per restriction R5. This creates several challenges in adapting DynamoRIO to run in SGX.

In the vanilla DynamoRIO design, the dynamic translation engine and the target application share the same thread, but they have separate TLS segment for cleaner context switch. DynamoRIO keeps the default TLS segment for the target application and creates a new TLS segment for itself at a different address. It switches between these 2 TLS segments by changing the segment register—DynamoRIO uses gsbase and the application uses `fsbase`.

**Multiplexing TLS Segments.** Normally, to context switch between threads, one TLS segment to save the currently executing application thread context is sufficient. But with Ratel, we need an additional TLS segment to save the state of Ratel itself. Furthermore, SGX itself reserves one additional TLS segment for its own internal use. This brings the total number of required TLS segments, for a correct context switch, to 3 on SGX.

But, the x86_64 architecture itself provides only 2 base registers (fsbase and gsbase) for storing pointers to TLS segments. Therefore, when we attempt to run DynamoRIO inside SGX, there are not enough base registers to save 3 TLS segment offsets (one each for DynamoRIO, SGX, and the application). We circumvent this limitation of the SGX platform as follows. First, Ratel adds 2 fields in each TLS segment to store fsbase and gsbase register values for that segment. We use these TLS segment fields to save and restore pointers to the segment base addresses. This allows us to still maintain and switch between 3 clean TLS segment views per thread. Second, when Ratel has to restore a TLS segment, it searches through a list of TLS segment base addresses, to find the right one to restore—this is because it does not have enough base registers to store 3 TLS segment bases (which would have avoided the search).

**Restoring TLS Segments on Context Switches.** Ratel conceptually maintains a linked list of a maximum of 3 TLS segment base pointers. The head of the list is the `fsbase` register, which serves as a pointer to the default first TLS segment created by SGX reserved for its own use. This default segment is called the *primary* and all TLS segments created subsequently are referred to as *secondary*. To point to the next TLS segment in the linked list, Ratel adds a new field in the TLS segment, which is NULL for the last element in the list. To traverse the list, the gsbase register is used. The list search begins with the primary, and the right segment to restore is always the last element in the link list. This way, Ratel can search and decide which of the 3 TLS segments to restore using only 2 registers (fsbase and gsbase).

There are two ways in which control can enter/exit from the enclave: via synchronous exits (e.g., ECALL/OCALL used for syscalls) and via asynchronous exits (e.g., used for exceptions, timer interrupts, and so on). During synchronous exits, Ratel sets up the TLS segment link list such that the restored TLS context state upon resumption is for DynamoRIO and SGX respectively. The subsequent exception handlers copy state from outside the enclave to inside, perform various Iago checks, and then setup the TLS segment to that of the application thread. During asynchronous exits, Ratel does not need to perform any special setup for the TLS segment link list. When the exception handler executes on resumption, it sets up the executing context just before the exit. Ratel performs similar checks and operations as in the case of synchronous exits, and then restores the context to that before the exit (which may be DynamoRIO's context or that of the application thread). Therefore, our design works correctly for both synchronous and asynchronous exits.

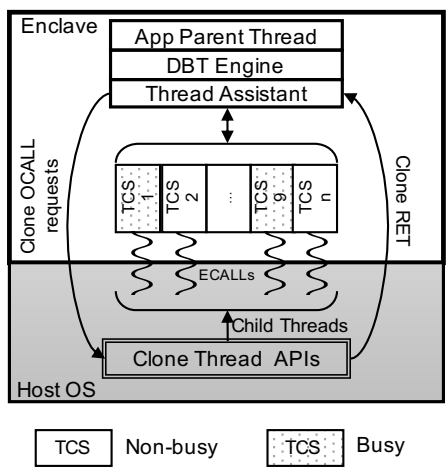

**Figure 3: Design for multi-threading in RATEL.**

**Dynamic Threading.** Since the number of TCS entries is fixed at enclave creation time on SGX, the maximum number of threads supported is capped. RATEL multiplexes the limited TCS entries available among the application threads dynamically, as shown in Figure 3. When an application wants to create a new thread (e.g., via `clone`), RATEL first checks if there is a free TCS slot. If it is the case, it performs an `OCALL` to do so outside the enclave. Otherwise, it busy-waits until a TCS slot is released. Once a TCS slot is available, the `OCALL` creates a new thread outside the enclave. After finishing thread creation, the parent thread returns back to the enclave and resumes execution. The child thread explicitly performs an `ECALL` to enter the enclave and DynamoRIO resumes execution for the application's child thread.

For all threading operations, RATEL ensures transparent context switches to preserve binary compatibility as intended by DynamoRIO. For security, RATEL creates and stores all thread-specific context either inside the enclave or SGX's secure hardware-backed storage at all times. It does not use any OS data structures or addresses for thread management.

### 4.4 Thread Synchronization

SGX provides basic synchronization primitives (e.g., SGX mutex and conditional locks) backed by hardware locks. But they can only be used for enclave code. Thus, they are semantically incompatible with the lock mechanisms used by DynamoRIO or legacy applications which use OS locks. For example, DynamoRIO implements a fast lock using the `futex` syscall, where the lock is kept in a shared memory accessible to all application threads and the OS. Here, the OS needs the ability to *read* the lock state to determine whether it should wait during the `FUTEX_WAIT` syscall.

A naive design would be to maintain the `futex` lock in public memory, such that it is accessible to the enclave(s) and the OS. However, the OS can arbitrarily change the lock state and attack the application. Specifically, it can reset the lock during the execution of a critical section in an application thread. Therefore, this design choice is not safe.

As an alternative, we can employ a two-copy mechanism for locks. The enclave can keep the lock in private memory. When it wants to communicate state change to the OS, RATEL can tunnel a `futex` state to the host OS. This approach is problematic as well. Threads inside the enclave may frequently update the locks in private memory. The futex state outside the enclave needs to be kept consistent with the private copy, when the OS kernel and the untrusted part of the enclave access it, or else the semantics of the lock may not uphold. The more frequent the local updates to the in-enclave copy of the lock state, the higher the chance of inconsistencies. In general, avoiding such race conditions usually involves using locks for synchronizing. But requiring locks to synchronize copies of other locks, as suggested in this design alternative, only results in a chicken-and-egg problem.

Supporting such semantics efficiently, where the OS has a shared read access to the lock state, is difficult with SGX because of restrictions $R1$ and $R3$. Figure 4 shows the schematics of design choices for implementing synchronization primitives available on SGX. Note that options (a) and (b) are insecure as discussed above. The option (c) is to keep the lock state private in a dedicated lock manager enclave to protect against OS attacks. This can be made conceptually secure, but is expensive because all lock state updates from application threads would convert into procedure calls to the dedicated lock manager enclave. RATEL uses the last alternative (d), which is also secure and expensive, but is a simpler version of (c) to implement. Specifically, instead of using a separate lock manager enclave, RATEL implements a lock manager inside the same enclave that executes the application. Our simplification has one limitation: Only threads within the same application process (and enclave) can utilize RATEL synchronization primitives.

RATEL **Lock Manager Implementation.** Our design choice of using a single enclave to execute both the DynamoRIO engine and all application threads eliminates considerable complexity in implementation. It turns out that a `futex`-based becomes unnecessary since no sharing across the process boundary or with the OS is needed. The DynamoRIO usage of futexes can thus be replaced with a simpler primitive such as spinlocks to achieve the same functionality. Specifically, RATEL implements a lock manager using the hardware spinlock exposed by SGX. RATEL invokes its in-enclave lock manager either when DynamoRIO uses futexes or when the application binaries perform lock-based synchronization. For the DynamoRIO code, we manually change it to invoke our lock manager. In case of application locks, RATEL loads application binary into the code cache and replaces thread-related calls (e.g., `pthread_cond_wait`) in the enclave-OS interface with stubs to invoke our lock manager to use RATEL-provided safe synchronization primitives.

### 4.5 Signal Handling

RATEL cannot piggyback on the existing signal handling mechanism exposed by the SGX, due to restriction $R5$. Specifically, when DynamoRIO executes inside the enclave, the DynamoRIO signal handler needs to get description of the event to handle it (Figure 5(a)). However, Intel's SGX platform software removes all such information when it delivers the signal to the enclave. This breaks the functionality of programmer-defined handlers to recover from well known exceptions (e.g., divide by zero). Further, any illegal instructions inside the enclave generate exceptions, which are raised in the form of signals. Existing binaries may not have handlers for recovering from such illegal instructions. Therefore, RATEL must provide handlers for all such exceptions.

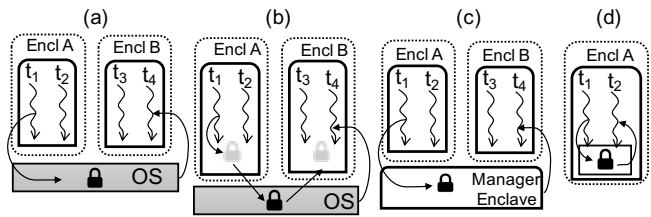

Figure 4: Lock synchronization design choices. (a) Futex. (b) Two-copy design with futex in public memory. (c) Dedicated lock manager in a separate enclave. (d) RATEL case: the threads and lock manager are in the same enclave.

Recall that SGX allows entering the enclave at fixed program points. Leveraging this, RATEL employs a *primary* signal handler that it registers with SGX. For any signals generated for DynamoRIO or the application, we always enter the enclave via the primary handler and we copy the signal number into the enclave. We then use the primary as a trampoline to route the control to the appropriate *secondary* signal handler inside the enclave, based on the signal number. At a high-level, we realize a virtualized trap-and-emulate signal handling design. We use SGX signal handling semantics for our primary. For the secondary, we setup and tear down a separate stack to mimic the semantics in the software. The intricate details of handling the stack state at the time of such context switch are elided here. Figure 5(b) shows a schematic of our design and we explain the flow of control (and associated issues) here.

**Registration.** The original DynamoRIO code and the application binary use `sigaction` to register signal handlers for itself. In RATEL, first we change DynamoRIO logic to register only the primary signal handler with SGX. We then record the DynamoRIO and application registrations as secondary handlers. This way, whenever SGX or the OS delivers the signal to the enclave, SGX directs the control to our primary handler[4]. Since this is a pre-registered handler, SGX allows it. The primary handler checks the signal information (e.g., signal code) and explicitly routes execution to the secondary.

**Delivery.** A signal may arrive when the execution control is inside the enclave. In this case, RATEL executes a primary signal handler that delivers the signal to the enclave. However, if the signal arrives when the CPU is in a non-enclave context, SGX does not automatically invoke the enclave to redirect execution flow. To force this, RATEL has to explicitly enter the enclave. But it can only enter at a pre-registered program point with a valid context, as per restriction *R5*. Thus, RATEL first wakes up the enclave at a valid point (via `ECALL`) and copies the signal information to private memory. It then simulates the signal delivery by setting up the enclave stack in private memory to execute the primary handler.

**Exit.** After executing their logic, handlers use `sigreturn` instruction for returning control to the point before the signal interrupted the execution. When RATEL observes this instruction in the secondary handler it has to simulate a return back to the primary handler instead. The primary handler then performs its own real `sigreturn`. SGX then resumes execution from the point before the signal was generated.

**Handling Nested Signals.** One issue with platforms like SGX is that it supports synchronous and asynchronous signals. Signals can be *nested*, in the sense that signals can be delivered by the OS while the enclave is handling another one. The enclave *cannot* mask signals selectively at runtime on SGX. Accordingly, the potential for subtle re-entrancy bugs in the enclave signal handling code arises. At a high-level, RATEL handles signals safely by ensuring that unsafe nesting is not possible. Specifically, the SGX platform automatically saves enclave state in private memory regions pointed to by hardware State Save Area (SSA) when an exception is to be delivered. To support nesting, SGX provides an array of SSAs frames, leaving the option to set the maximum size of array (hence, the maximum possible nesting depth) to the enclave. RATEL utilizes this feature to limit the nesting depth to 2. This is needed because in RATEL one SSA frame can be used by RATEL itself and the other by the target application. With the maximum nesting depth set to 2, if a nested signal of depth 3 is being attempted to be delivered, SGX securely aborts the enclave since not enough SSA frames are available. With this design, there are only 4 possible combinations of re-entrancy to reason about:

(1) RATEL signal handler is interrupted with a signal to be delivered to the application;

(2) application's signal handler is interrupted with a signal to be delivered to RATEL;

(3) RATEL signal handler is interrupted with a signal to be delivered to the RATEL;

(4) application handler is interrupted with a signal to be delivered to the application;

In Case 1 and 3, DynamoRIO gets execution control. In Case 1, DynamoRIO has one additional SSA frame available to save the current state of the primary handler and deliver control to the beginning of the primary signal handler but with the new context. In Case 3, the 2 SSA frames are already used up to save RATEL's primary handler state and the application's secondary handler state. Therefore, SGX is unable to deliver the signal and it is ignored at this point in time. Both these scenarios are safe from re-entrancy bugs. In Case 2 and Case 4, similar to Case 3, the 2 SSA frames are already in use to store the current state of the primary and secondary signal handlers respectively. SGX, thus, cannot deliver the signal and it is ignored, which is safe. Thus, in all the above scenarios, RATEL design handles reentrancy safely. Note that RATEL reentrancy handling for synchronous exceptions, which are used for threads and syscalls (Section 4.3), and asynchronous exceptions is the same regardless of the exception type.

*Remark.* RATEL signal handling extends the trap-and-emulate principle used by DynamoRIO. Since DynamoRIO design enforces transparency, we preserve this in RATEL. This goes beyond merely working around the SGX limitations, thus making our design different than existing frameworks. Specifically, existing library OS-based SGX frameworks (e.g., Graphene-SGX, Occlum) assume that all exception registration and execution will go via the prescribed library interfaces. So, they do not keep a separate signal context for the library OS signals and the application. These works do not discuss what happens during nested signals or when the enclave-OS interaction happens outside of the prescribed library registration and handler mechanisms.

---

[4]Vanilla SGX PSW does not provide an API which allows the enclave to register for signals, we have changed the PSW to support this.

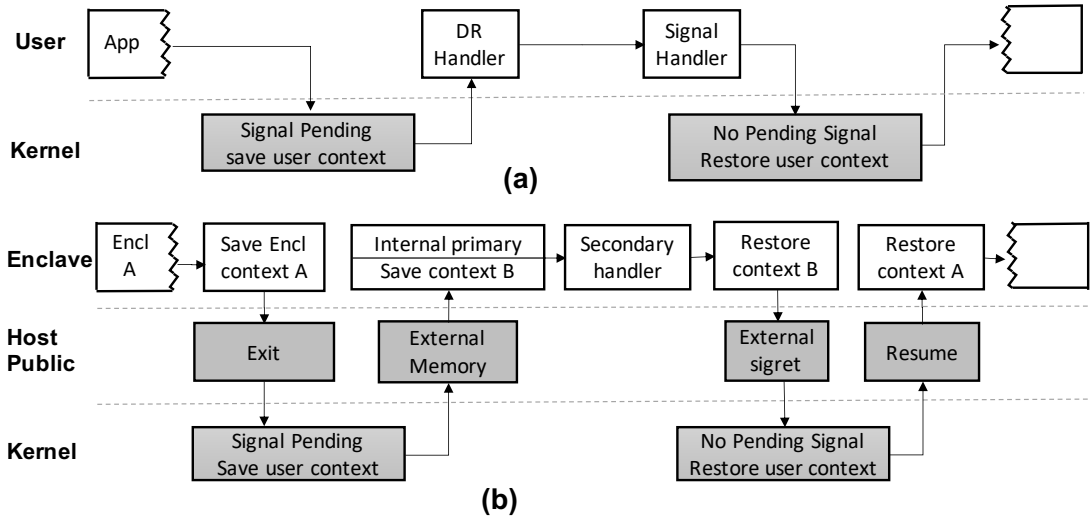

**Figure 5: (a) Original signal handling in DynamoRIO. (b) Signal handling in Ratel.**

## 5 IMPLEMENTATION

We implement Ratel on DynamoRIO [23]. We run DynamoRIO inside an enclave with the help of standard Intel SGX development kit that includes user-land software (SDK v2.1), platform software (PSW v2.1), and a Linux kernel driver (v1.5). We make a total of 9667 LoC software changes to DynamoRIO and SDK infrastructure. We run Ratel on an unmodified hardware that supports SGX v1.

Ratel design makes several changes to DynamoRIO core (e.g., memory management, lock manager, signal forwarding). We discuss three high-level implementation challenges which arise in implementing our design, while eliding lower-level challenges here for brevity. The root cause of our highlighted challenges is the way Intel SDK and PSW expose hardware features and what DynamoRIO expects.

**Self-identifying Load Address.** The vanilla DynamoRIO engine needs to know its own start location in memory to avoid overlapping its own address space with that of the target application, and so it uses a hard-coded address. Since our modifications change such hard-coded address assumptions, Ratel uses a `call – pop` instruction sequence to self-identify the runtime location in memory for the DynamoRIO engine [67, 85], aligns it at a page boundary, and updates the DynamoRIO logic to use code location-independent addressing.

**Setting SSA Slots.** The vanilla SGX SDK and PSW use just two SSA frames: one is used to process timer interrupts specially and the other to handle all other interrupts. As explained in Section 4.5, the Ratel design aims to set the effective nesting depth to 2. Therefore, in the implementation, it uses 3 SSA slots: one reserved for the timer interrupt (to be handled by the SGX SDK), and the remaining two as described in Section 4.5. The timer interrupt handler simply routes control to whichever signal handler was interrupted. The SGX specification allows setting the required SSA slots by changing the NSSA field in our SDK implementation.

**Preserving Execution Contexts.** For starting execution of a newly created thread, Ratel invokes a pre-declared ECALL to enter the enclave. This is a nested ECALL, which is not supported by SGX SDK. To allow it, we modify the SDK to facilitate the entrance of child threads and initialize the thread data structure for it. Specifically, we check if the copy of thread arguments inside the enclave matches the ones outside before resuming thread execution. We save specific registers so that the thread can exit the enclave later. Note that the child thread has its own execution path differentiating from the parent one, Ratel hence bridges its return address to the point in the code cache that a new thread always starts. After the thread is initialized, we explicitly update DynamoRIO data structures to record the new thread (e.g., the TLS base for application libraries) This way, DynamoRIO is aware of the new thread and can control its execution in the code cache.

**Propagating Implicit Changes & Metadata.** Thread uses exit/exit_group syscall for terminating itself. Then the OS zeros out the child thread ID (`ctid`). In Ratel, we explicitly create a new thread inside the enclave, so we have to terminate it explicitly by zeroing out the pointers to the IDs. Further, we clean up and free the memory associated with each thread inside and outside the enclave.

**Built-in Profilers.** DynamoRIO supports two modes of instrumentation — built-in profilers and client plugins. Profilers are readily available with the DynamoRIO core and provide basic functionalities such as instruction tracing, logging, and tuning the DynamoRIO parameters (see Table 10). On the other hand, DynamoRIO clients are specific instrumentation plugins designed to perform user-desired tasks (e.g., Shadow stack). Since profilers are part of DynamoRIO, Ratel supports all of them out-of-the-box. Our current implementation removes client plugin support to reduce TCB. We demonstrate Ratel compatibility with these profilers in Section 6.4.

| Subsystem | Total | Impl | Implementation | | | Covered |
|---|---|---|---|---|---|---|
| | | | Del | Emu | P.Emu | DR + Binaries |
| Process | 12 | 8 | 4 | 2 | 2 | 3 |
| Filename based | 37 | 25 | 25 | 0 | 0 | 16 |
| Signals | 12 | 7 | 3 | 4 | 0 | 6 |
| Memory | 18 | 10 | 6 | 0 | 4 | 4 |
| Inter process communication | 12 | 4 | 4 | 0 | 0 | 0 |
| File descriptor based | 65 | 53 | 48 | 0 | 5 | 30 |
| File name or descriptor based | 19 | 9 | 9 | 0 | 0 | 5 |
| Networks | 19 | 17 | 15 | 0 | 2 | 15 |
| Misc | 124 | 79 | 79 | 0 | 0 | 36 |
| Total | 318 | 212 | 193 | 6 | 13 | 115 |

**Table 2:** Ratel **syscall support. Column** $2 - 3$**: total Linux system calls and support in** Ratel**. Column** $4 - 6$**: syscalls implemented by full delegation, full emulation, and partial emulation respectively. Column** 7**: syscalls tested in** Ratel**.**

## 6 EVALUATION

We evaluate the following properties of Ratel empirically:

- *Binary compatibility.* How well does Ratel provide binary compatibility with common Linux programs on SGX?
- *TCB.* What is the size of the trusted computing base (TCB) for Ratel?
- *Performance.* How much overhead does Ratel introduce for applications?
- *Graphene-SGX.* Does Ratel compare in compatibility and performance to Graphene-SGX, the state-of-the-art library OS for SGX.
- *Instrumentation capability.* What kinds of low-level monitors does Ratel provide for end applications?

**Setup.** All our experiments are performed on a Lenovo machine with SGX v1 support, 128 MB EPC of which approximately 90 MB is available for user-enclaves, 12 GB RAM, 64 KB L1, 256 KB L2, 4096 KB L3 cache, 3.4GHz processor speed. We use Ubuntu 16.04, Intel SGX SDK v2.1, PSW v2.1, driver v1.5, DynamoRIO v6.2.17. All performance statistics reported are the geometric mean over 5 runs. To foster open science and reproducibility, we have made our implementation and evaluation public. Ratel code-base, our benchmarks, and case-studies are open-source and available [1].

To compare Ratel's binary compatibility and performance with other approaches, we have chosen Graphene-SGX, a library OS which runs inside SGX enclave. Graphene-SGX offers the lowest compatibility barrier of all prior systems to our knowledge, specifically offering compatibility with glibc. It is a mature and a publicly available system, which has been maintained for over 3 years as of this writing.

## 6.1 Compatibility

To evaluate compatibility, we initially select 310 binaries that cover an extensive set of benchmarks, utilities, and large-scale applications. These are commonly reported to be used as evaluations target for DynamoRIO and prior enclave-based systems [15, 24, 28, 32, 44, 73, 74] that we surveyed for our study. Further, they represent a mix of memory-intensive, CPU-intensive, multi-threading, network-intensive, and file I/O workloads. A total of 69 binaries are from micro-benchmarks: 29 from SPEC 2006 (CPU), 1 from IOZone (I/O) v3.487, 9 from FSCQ v1.0 (file API), 21 from HBenchOS v1.0 (system stress-test), and 9 from Parsec-SPLASH2 (multi-threading). We run 12 binaries from 3 real-world applications—cURL v7.65.0 (server-side utilities), SQLite v3.28.0 (database), Memcached v1.5.20 (key-value store), and 9 applications from Privado (secure ML framework). We selected all 229 Linux utilities which are available from our test system's /bin and /usr/bin directories. Apart from these 310 binaries, we tested 2 language runtimes, Python (v2.7.17 and v3.8.2) and R v3.6.3. Python is tested with Python-CPU-Benchmark [6], Python Programming Examples [8], and PyBench v2.0 [7]. R is tested with the R-benchmark-25 [3].

### 6.1.1 Compatibility Gains.
Python and R benchmarks run out-of-the-box on Ratel, highlighting how multiple languages can be supported without any special handling. The R interpreter has JIT enabled, which constitutes an example of how Ratel can handle dynamically generated code gracefully, whereas the Python interpreter is bytecode-interpretation based. For the remaining 310 benchmarks and applications, we download the source code and compile it with default flags required to run them natively on our machine. We directly use the existing binaries for Linux utilities. We test the same binaries on native hardware, with DynamoRIO, and with Ratel without changing the original source-code or the binaries. Out of 310 binaries, 272 of targets execute successfully with the native Linux and with the (unmodified) vanilla DynamoRIO. The remaining 38 binaries either use unsupported devices (e.g., NTFS) or do not run on our machine. So we discard them from our Ratel experiments, since vanilla DynamoRIO also does not work on them. Of the remaining 272 binaries that work on the baselines, Ratel has support for the system calls used by 208 of these. Ratel runs these out-of-the-box with no additional porting effort.

**System Call Support & Coverage.** Ratel supports a total of 212/318 (66.66%) syscalls exposed by the Linux Kernel. We emulate 6 syscalls purely inside the enclave and delegate 193 of them via OCALLs. For the remaining 13, we use partial emulation and partial delegation. Table 2 gives a detailed breakdown of our syscall support. Syscall usage is not uniform across frequently used applications and libraries [78]. Hence we empirically evaluate the degree of expressiveness supported by Ratel. For all of the 272 binaries in our evaluation, we observe a total of 121 unique syscalls are used by the benchmarks. Ratel supports 115 of them. Table 2 shows the syscalls supported by Ratel and their usage in our benchmarks and real-world applications (See Section 6.1.2). Figure 6a and 6b show the distribution of unique syscalls and their frequency as observed over binaries supported by Ratel. Thus, our empirical study shows that Ratel supports 115/121 (95.0%) syscall observed in our benchmark programs.

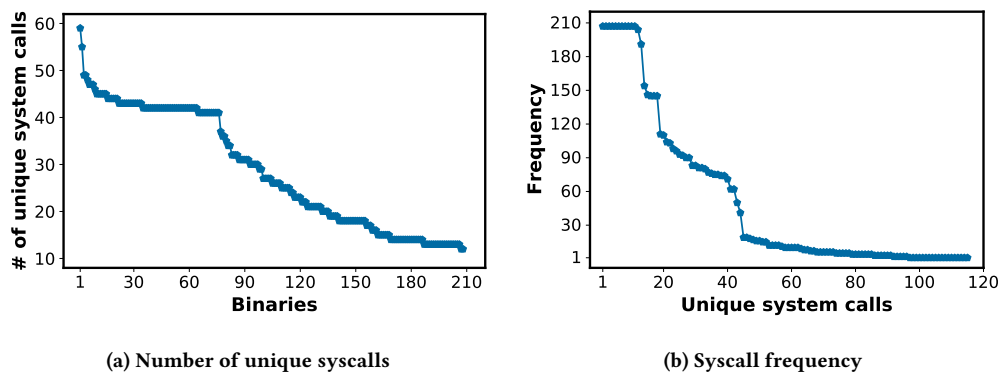

**(a) Number of unique syscalls**

**(b) Syscall frequency**

**Figure 6: System calls statistics over all 208 binaries. (a) Unique syscalls for each binary; and (b) frequency per syscall.**

To support 212 syscalls, we added 3233 LoC (15 LoC per syscall on average). In the future, RATEL can be extended to increase the number of supported syscalls. Readers are referred to Section 6.1.2 for more details. RATEL handles 31 out of the 32 standard Linux signals—the SIGPROF signal is not handled which vanilla DynamoRIO itself does not support.

**Library vs Binary Compatibility.** We maintain full binary compatibility with all 208 binaries tested for which we had system call support in RATEL. For them, RATEL works out-of-the-box in our experiments. We report that, given the same inputs as native execution, RATEL produces same outputs. The advantage of RATEL is that it makes no assumptions about which specific implementation or version of libc or higher-level API the application uses. To validate that this assumption is indeed empirically preserved, we test RATEL with binaries that use different libc implementations. Specifically, we compile HBenchOS benchmark (12 binaries), which is a OS system stress testing benchmark, with two different libc versions: glibc v2.23 and musl libc v1.2.0. We report that RATEL executes these benchmarks out-of-the-box with both the libraries, without any modification or specialization to RATEL implementation.

Lastly, as a point of comparison, we report our experience on porting our micro-benchmarks to the state-of-the-art library-compatibility system for SGX (Graphene-SGX) in Section 6.1.3. Of the 75 programs tested, Graphene-SGX fails on 13. RATEL works correctly for all except 1, which failed only due to the virtual memory limits of SGX hardware.

### 6.1.2 Detailed Breakdown of Compatibility Tests.

We provide detailed breakdown of the compatibility observed for our Linux utilities and other benchmarks.

**Linux Utilities.** Our tests include all the Linux built-in binaries available on our experimental Ubuntu system. These comprise 229 shared-objected binaries in total, which are typically are in the directories /bin and /usr/bin.

We run each utility with the options and inputs, representative of their common usage. The specifics of the input configurations are reported as a script in our released system publicly. Out of 229 benchmarked utilities, 195 worked with our test machine natively and with vanilla DynamoRIO. Of these 195 binaries, a total of 138 have all system calls presently supported in RATEL, all of which worked correctly in our tests out-of-the-box. The 57 programs that did not work fail for 2 reasons: missing syscall support and virtual memory limits imposed by SGX. Table 3 and Table 4 list all Linux utilities and binaries from real-world applications and benchmarks that ran successfully, and present the number of unique system calls for each. Table 5 and Table 6 summarize the reasons for all binaries that fail in RATEL and in native and DynamoRIO, respectively.

For the incompatible cases, 45 fail due to lack of multi-processing (fork) support in RATEL. As explained in Section 3.3, not supporting fork is an explicit design decision in RATEL. 5 utilities use certain POSIX signals, which are outside the 32 standard signals in POSIX, for which presently RATEL has incomplete support (e.g., real-time signals SIGRTMIN + n). Another 5 utilities fail because they invoke other system calls which the restriction R3 in SGX fundamentally does not permit (e.g., shared memory syscalls such as shmat, shmdt, shmctl, etc.). These syscalls are not supported in RATEL.[5] 1 utility which fails is because of the virtual memory limit in SGX, as it loads more than 100 shared libraries. The remaining 1 utility fails because RATEL has no support to execv syscall.

**Other Benchmarks & Applications.** From the 81 binaries from micro-benchmarks and real applications, 11 do not work with RATEL. 5 binaries from HBenchOS (lat_proc, lat_pipe, lat_ctx, lat_ctx2, bw_pipe) either use fork or shared memory system calls disallowed by R3. 2 binaries (lat_memsize from HBenchOS, mcf from SPEC 2006) with DynamoRIO require virtual memory larger than SGX limits on our experimental setup. The remainder (e.g., bwaves from SPEC 2006) fail to run even on the baseline setup of our Linux OS with vanilla DynamoRIO.

### 6.1.3 Comparison to Graphene-SGX.

Applications using Graphene-SGX have to work only with a specific library interface, namely a custom glibc, which requires re-linking and build process changes. RATEL, in contrast, has been designed for binary compatibility which is a fundamental difference in design. To demonstrate the practical difference, we reported in Section 6.1 that HBenchOS benchmark works out-of-the-box when built with both glibc and musl, as an example.

---

[5]Note that the ioctl syscall involves more than 100 variable parameters. RATEL syscall stubs currently does not cover all of them.

| Utility | # of sys. | Utility | # of sys. | Utility | # of sys. | Utility | # of sys. | Utility | # of sys. | Utility | # of sys. |
|---|---|---|---|---|---|---|---|---|---|---|---|
| ed | 18 | ppdpo | 29 | dirmngr | 21 | hcitool | 20 | systemctl | 32 | systemd-cgtop | 25 |
| cvt | 13 | psnup | 14 | enchant | 20 | bluemoon | 21 | vim.basic | 45 | dirmngr-client | 13 |
| eqn | 36 | t1asm | 13 | epsffit | 13 | btattach | 21 | hciconfig | 25 | systemd-escape | 18 |
| gtf | 18 | troff | 15 | faillog | 18 | fwupdate | 19 | brltty-ctb | 21 | systemd-notify | 19 |
| pic | 13 | uconv | 13 | gendict | 14 | gatttool | 24 | fusermount | 19 | wpa_passphrase | 18 |
| tbl | 13 | bccmd | 25 | hex2hcd | 18 | gencnval | 13 | journalctl | 26 | gamma4scanimage | 13 |
| xxd | 14 | btmgmt | 24 | icuinfo | 15 | lessecho | 13 | sudoreplay | 16 | systemd-analyze | 31 |
| curl | 32 | busctl | 37 | kbxutil | 14 | loginctl | 31 | watchgnupg | 18 | systemd-inhibit | 31 |
| derb | 23 | catman | 13 | lastlog | 14 | makeconv | 14 | xmlcatalog | 12 | systemd-resolve | 31 |
| find | 27 | cd-it8 | 21 | lesskey | 13 | ppdmerge | 26 | zlib-flate | 13 | ulockmgr_server | 18 |
| gawk | 25 | expiry | 16 | lexgrog | 15 | psresize | 14 | cupstestdsc | 26 | systemd-tmpfiles | 34 |
| grep | 21 | genbrk | 14 | manpath | 14 | psselect | 14 | cupstestppd | 18 | gpg-connect-agent | 22 |
| htop | 26 | gencfu | 13 | obexctl | 35 | t1binary | 13 | hostnamectl | 30 | kerneloops-submit | 20 |
| kmod | 17 | grotty | 13 | pkgdata | 14 | t1binary | 13 | systemd-run | 29 | evince-thumbnailer | 21 |
| ppdc | 30 | l2ping | 21 | ppdhtml | 27 | t1disasm | 13 | timedatectl | 32 | fcitx-dbus-watcher | 18 |
| ppdi | 30 | l2test | 27 | preconv | 15 | t1disasm | 13 | brltty-trtxt | 22 | systemd-detect-virt | 18 |
| qpdf | 14 | psbook | 13 | sdptool | 22 | transfig | 15 | dbus-monitor | 31 | dbus-cleanup-sockets | 19 |
| gpg2 | 27 | pstops | 14 | ssh-add | 20 | vim.tiny | 34 | dbus-uuidgen | 13 | systemd-stdio-bridge | 25 |
| wget | 29 | rctest | 25 | t1ascii | 13 | dbus-send | 30 | fcitx-remote | 23 | systemd-ask-password | 20 |
| btmon | 23 | soelim | 12 | udevadm | 27 | gpg-agent | 18 | gpgparsemail | 14 | webapp-container-hook | 26 |
| genrb | 13 | whatis | 20 | volname | 15 | hciattach | 18 | systemd-hwdb | 17 | systemd-machine-id-setup | 18 |
| grops | 14 | rfcomm | 19 | xmllint | 15 | localectl | 30 | systemd-path | 18 | systemd-tty-ask-password-agent | 25 |
| mandb | 27 | bootctl | 19 | ciptool | 20 | pg_config | 16 | enchant-lsmod | 13 | systemd-update-activation-environment | 31 |

**Table 3: List of GNU utilities (138) tested with Ratel and the corresponding the number of unique system calls invoked in their single execution.**

| Utility | # of sys. | Utility | # of sys. | Utility | # of sys. | Utility | # of sys. | Utility | # of sys. | Utility | # of sys. |
|---|---|---|---|---|---|---|---|---|---|---|---|
| gcc | 43 | dealII | 43 | leslie3d | 43 | xalancbmk | 49 | bw_mmap_rd | 42 | water_spatial | 44 |
| fmm | 45 | soplex | 43 | calculix | 43 | LFS-write | 43 | resnet50app | 41 | inceptionv3app | 42 |
| curl | 55 | povray | 43 | GemsFDTD | 42 | multiopen | 41 | densenetapp | 42 | multicreatemany | 41 |
| milc | 42 | barnes | 45 | specrand | 42 | multiread | 41 | multicreate | 47 | multicreatewrite | 41 |
| namd | 42 | iozone | 47 | specrand | 42 | memcached | 59 | lat_fslayer | 42 | lat_syscall(sbrk) | 42 |
| bzip2 | 48 | lat_fs | 41 | lenetapp | 41 | radiosity | 44 | lat_connect | 42 | lat_syscall(write) | 42 |
| gobmk | 42 | bw_tcp | 44 | vgg19app | 43 | ocean_ncp | 44 | resnext29app | 42 | lat_syscall(getpid) | 42 |
| hmmer | 42 | h264ref | 42 | raytrace | 45 | bw_mem_cp | 42 | resnet110app | 42 | lat_syscall(sigaction) | 42 |
| sjeng | 42 | omnetpp | 43 | ocean_cp | 45 | bw_mem_rd | 42 | widaresnetapp | 43 | lat_syscall(getrusage) | 43 |
| tonto | 44 | gromacs | 45 | volerand | 45 | bw_mem_wr | 42 | squeezenetapp | 41 | lat_syscall(gettimeofday) | 42 |
| astar | 43 | lat_sig | 44 | bw_bzero | 42 | libquantum | 42 | LFS-smallfile | 49 | | |
| sqlite | 47 | lat_tcp | 42 | lat_mmap | 42 | multiwrite | 41 | LFS-largefile | 42 | | |
| zeusmp | 43 | lat_udp | 42 | cactusADM | 43 | bw_file_rd | 42 | water_nsquare | 44 | | |

**Table 4: List of applications (12) and individual benchmarks (63) tested with Ratel and the corresponding number of unique system calls invoked in their single execution.**

| Reason category | # unsuc. | Case examples |
|---|---|---|
| fork | 49 | strace, scp, lat_proc and lat_pipe from HBenchOS, etc. |
| execv | 1 | systemd-cat |
| signal | 5 | colormgr, cd-iccdump, bluetoothctl etc. |
| Unsupported syscalls | 6 | webapp-container, webbrowser-app, etc. |
| Out-of-memory | 3 | shotwell, mcf from SPEC 2006, lat_memsize from HBenchOS |

**Table 5: Summary of the reasons for failure of all 64 unsuccessful binaries tested with Ratel.**

| Reason category | # unsuc. | Case examples |
|---|---|---|
| NTFS related | 16 | ntfs-3g, ntfs-3g.probe, ntfs-3g.secaudit, etc. |
| Printer related | 7 | lp, lpoptions, lpq, lpr, lprm, etc. |
| Scanner related | 2 | sane-find-scanner, scanimage |
| Failure in native run | 5 | umax_pp, cd-create-profile, and bwaves from SPEC 2006, etc. |
| Failure in DynamoRIO run | 8 | ssh, ssh-keygen, dig, etc. |

**Table 6: Summary of the reasons for failure of all 38 unsuccessful binaries tested with Linux and DynamoRIO.**

Graphene-SGX requires a manifest file, for each application, that specifies the main binary name as well as dynamic-libraries, directories, and files used by the application. By default, Graphene-SGX does not allow creation of new files during runtime. We use the `allow_file_creation` to disable this default. We tested all 75 benchmark and application binaries (HBenchOS, Parsec-SPLASH2, SPEC, IOZone, FSCQ, SQLite, cURL, Memcached, Privado), out of which 62 work with Graphene-SGX. *Of the 13 that fail on Graphene-SGX, all except 1 work on Ratel, with the only failure being due to virtual memory limits.*

For Graphene-SGX, 3/9 Parsec-SPLASH2 binaries (water_nsquare, water_spatial and volrend), IOZone binary, and SQLite database workload [53] failed due to I/O error, (e.g., [37]) which is an open issue. 3/24 binaries from SPEC 2006 failed. Graphene-SGX fails for cactusAMD due failed due to a signal failure, which is mentioned as an existing open issue on its public project page [14]. The calculix program fails with a segmentation fault. The omnetpp could not process the input file in-spite of making the input file as allowed in the corresponding manifest file. 4 networking related binaries from HBenchOS namely lat_connect, lat_tcp, lat_udp and bw_tcp could not run, resulting in a "bad address" error while connecting to localhost. lat_memsize from HBenchOS fails on Graphene-SGX as it fails on RATEL too due to the virtual memory limit.

| Function | Trusted | | | Untrusted | | | | |
|---|---|---|---|---|---|---|---|---|
| | SDK+PSW | DR | Total | SDK+PSW | DR | Driver | Host | Total |
| Original | 147928 | 129875 | 277803 | 49838 | 66629 | 2880 | 1769 | 121116 |
| Loader | 69 | 1604 | 1673 | 27 | 89 | N/A | 332 | 448 |
| MM | 46 | 2241 | 2287 | 44 | 0 | N/A | 0 | 44 |
| Syscalls | 0 | 1801 | 1801 | 0 | 0 | N/A | 1432 | 1432 |
| Instr | 0 | 45 | 45 | 0 | 0 | N/A | 26 | 26 |
| TLS | 18 | 60 | 78 | 18 | 0 | N/A | 0 | 18 |
| Signals | 201 | 236 | 437 | 136 | 0 | N/A | 0 | 136 |
| Threading | 389 | 393 | 782 | 130 | 0 | N/A | 157 | 287 |
| Sync | 0 | 173 | 173 | 0 | 0 | N/A | 0 | 0 |

**Table 7: Breakdown of RATEL TCB.**

## 6.2 TCB Breakdown

We trust Intel SGX support software (SDK and PSW) that executes inside the enclave and interfaces with the hardware. This choice is same as any other system that uses enclaves. RATEL comprises one additional trusted component—DynamoRIO. Put together, RATEL amounts to 277,803 LoC TCB. This is comparable to existing SGX frameworks that have 100K to 1M LoC [15, 28], but provide library-based compatibility at best.

Table 7 (columns 2-3) summarizes the breakdown of the LoC included in the trusted components of the PSW, the SGX SDK, the DynamoRIO system, as well as the code contributed by each of the sub-systems supported by RATEL. The original DynamoRIO engine comprises 353, 139 LoC. We reduce it to 129, 875 LoC (trusted) and 66, 629 LoC (untrusted) by removing the components that are not required or used by RATEL. Then we add 8, 589 LoC to adapt DynamoRIO to SGX as per the design outlined in Section 4. Apart from this, as described in Section 5, we change the libraries provided by Intel SGX (SDK and PSW) and add 1, 078 LoC.

Of the 277,803 LoC of trusted code, 123,322 LoC is from the original DynamoRIO code base responsible for loading the binaries, code cache management, and syscall handling. 110,848 LoC and 37,080 LoC are from Intel SGX SDK and PSW respectively. RATEL implementation adds only 6,553 LoC on top of this implementation. A large fraction of our added TCB (27.5%) is because of the OCALL wrappers that are amenable to automated testing and verification [48, 74]. Rest of the 4,752 LoC are for memory management, handling signals, TLS, and multi-threading interface.

RATEL relies on, but does not trust, the code executing outside the enclave in the host process (e.g., OCALLs). This includes 2,391 LoC changes. We give a detailed breakdown of this in Table 7 (Columns 5-9).

## 6.3 Performance Analysis

We present the performance implications of our design choices made in RATEL. We have two main findings. First, the performance overheads vary significantly based on the application workload. Second, most of the overheads come from SGX restrictions R1-R5 and the enclave physical memory limits specific to our present test hardware. We point out that future SGX implementations may have over a 1000× larger private physical memory (1 TB EPC) compared to our test system [4]. Therefore, we expect that the performance bottlenecks due to physical memory limitations can be eliminated and are not fundamental to RATEL design. For completeness, we report RATEL memory footprint and impact of 90 MB limit in Section 6.3.2.

**Methodology for Performance Measurement.** For each target binary, we record the execution time in 3 settings:

- Baseline 1 (Linux). We execute the application binary directly with the native Linux (without SGX and DynamoRIO).
- Baseline 2 (DynamoRIO). We execute the application binary directly with DynamoRIO (without SGX) on Linux.
- RATEL. We use RATEL to execute the application binary in the enclave. We offset the execution time by deducting the overhead to create, initialize, load DynamoRIO and the application binary inside the enclave, and to destroy the enclave. It is well-known that SGX incurs a high overhead for enclave creation and attestation. However this is a one-time cost per application. Server-end applications, as studied in this work, have long execution time and can tolerate high initialization time. To avoid skewing the performance overheads, we deduct the enclave setup and tear-down overheads. This allows us to present a fair comparison of the actual execution overheads of RATEL with respect to Linux and DynamoRIO. Several previous works adopt the same measurement setup [19, 72].

To measure performance overheads, we collect various statistics of the execution profile of 58 program in our micro-benchmarks and 4 real-world applications (12 binaries in total). Specifically, we log the target application LoC, binary size, number of OCALLs, ECALLs, syscalls, enclave memory size, peak virtual memory (VmPeak) for Linux and DynamoRIO, untrusted and trusted VmPeak for RATEL, number of page faults, and number of context switches. We refer readers to Appendix A.1 and A.2 for detailed performance breakdowns. Table 9 provides detailed statistics. We also provide the overheads comparison between our Baseline 2 (DynamoRIO) and Baseline 1 (Linux) in all the related tables and figures to provide a breakdown of the performance overhead.

### 6.3.1 Performance Breakdown.

There are two main avenues of overhead costs we observe.

First, fundamental limitations of SGX result in increased memory-to-memory operations (e.g., two-copy design) or usage of slower constructs (e.g., spin-locks instead of fast futexes). Our evaluation on system stress workloads for each subsystem measure the worst-case cost of these operations. We report that on an average, SPEC CPU benchmarks result in 217.80% and 34.91% overheads (Figure 7), when compared to vanilla Linux (without DynamoRIO or SGX) and to DynamoRIO (without SGX) baselines respectively, while I/O-intensive workloads cost 87.5% and 66.2% slowdown (Figure 8 for IOZone benchmarks). Further, the performance overheads increases with larger I/O record sizes. The same is observed for HBenchOS binaries as reported in Table 8. The expensive spin-locks incur cost that increases with number of threads (Figure 9 for Parsec-SPLASH2 benchmarks). Overall, we observe that benchmarks that require large memory copies consistently exhibit significant slowdowns compared to others, highlighting the costs imposed by the two-copy design. The cost of signal handling also increases due to added context saves and restores in Ratel, as seen in a dedicated benchmark of HBenchOS (see last two rows in Table 8).

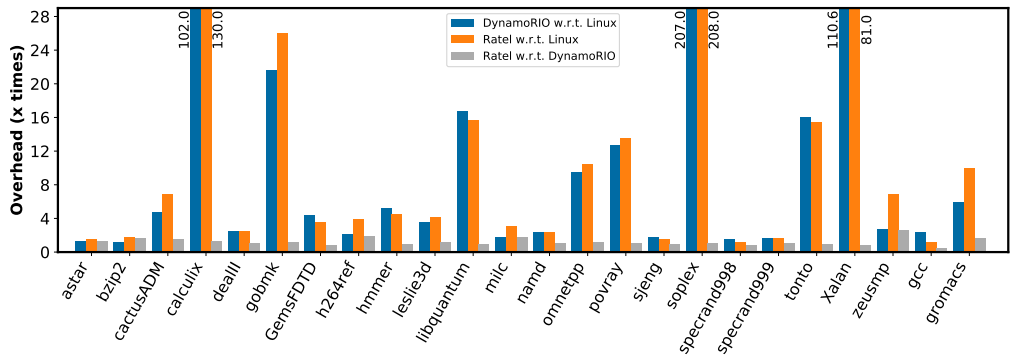

**Figure 7:** Ratel **performance for SPEC 2006 (CPU). Vanilla DynamoRIO execution time w.r.t. Linux,** Ratel **execution time w.r.t. Linux, and** Ratel **execution time w.r.t. vanilla DynamoRIO; lower value of overheads indicates better performance.**

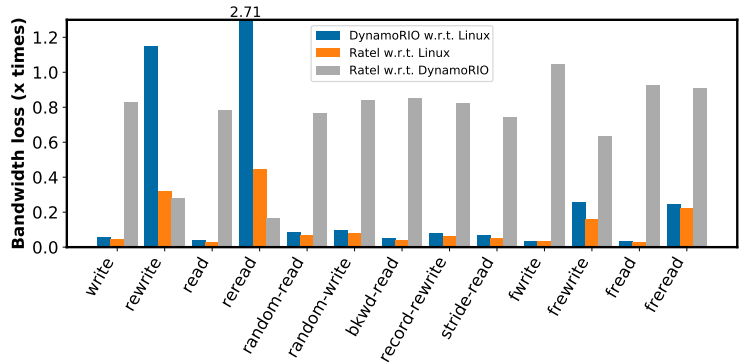

**Figure 8:** Ratel **performance for IOZone. Vanilla DynamoRIO bandwidth w.r.t. Linux,** Ratel **bandwidth w.r.t. Linux, and** Ratel **bandwidth w.r.t. vanilla DynamoRIO; value close to** 0 **indicates smaller bandwidth loss and hence better performance.**

Second, the current SGX hardware implementation has limited secure physical memory (called the EPC) of 90 MB. Executing anything on a severely limited memory resource results in large slowdowns (e.g., increased page-faults). Further, cost of each page-in and page-out operation itself is higher in SGX because of hardware based memory encryption. We measure the impact of this limitation by executing benchmarks and applications that exceed the working set size of 90 MB for both data and code. For example, we test varying download sizes in cURL (Figure 11a) and database sizes in SQLite (Figure 10a). When the data exceeds 90 MB, we observe a sharp increase in throughput loss. Similarly, when we execute varying sizes of ML models that require increasing size of code page memory, we observe increase in page faults and lowered performance (Figure 11b). We observe similar loss of latency and throughput when applications reach a critical point in memory usage as in memcached (Figure 10b). Appendix A.1 and A.2 detail the performance breakdown. The memory footprint of our system is reported separately in Section 6.3.2 as well.

**Performance Comparison with Graphene-SGX.** The performance overheads in Ratel vary based on workloads. This is observed for Graphene-SGX as well. As a direct point of comparison, we tested HBenchOS—a benchmark with varying workloads—with Graphene-SGX and find similar variation in performance based on the workload. The performance overheads of Graphene-SGX for HBenchOS benchmarks as compared to Linux, DynamoRIO, and Ratel is reported in Table 8. The slowdown in both Ratel and Graphene-SGX systems is comparable for I/O benchmarks, since both of them incur two copies. Graphene-SGX is significantly faster than the DynamoRIO baseline and Ratel for syscall and signal handling, because it implements a library OS inside the enclave and avoids expensive context switches. Ratel delegates most of the system calls to the OS and does not emulate it like Graphene-SGX, offering compatibility with multiple libraries in contrast. Further,

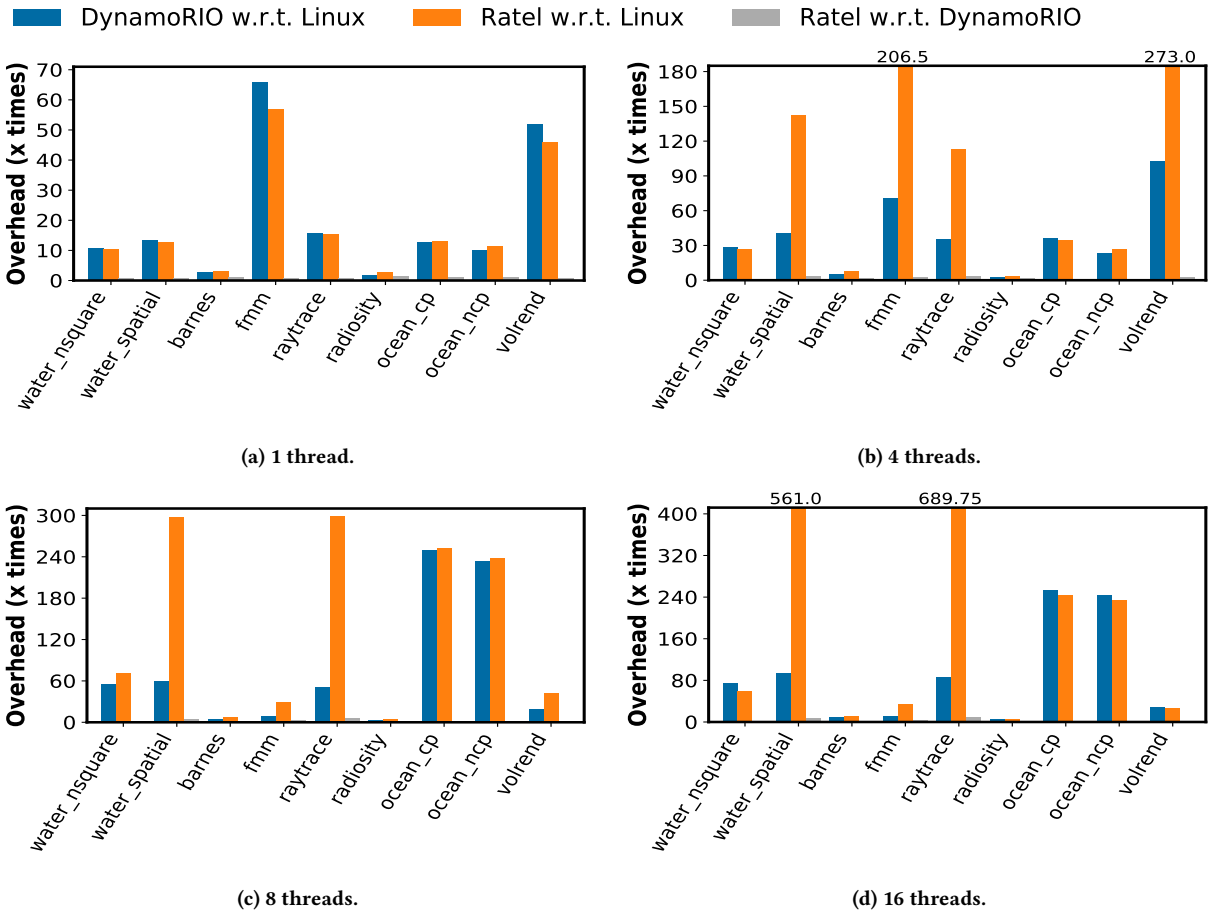

**Figure 9:** RATEL performance for Parsec-SPLASH-2 (multi-threading): (a), (b), (c), and (d) shows vanilla DynamoRIO execution time overhead w.r.t. Linux, RATEL execution time overhead w.r.t. Linux, and RATEL execution time overhead w.r.t. vanilla DynamoRIO, with 1, 4, 8, and 16 thread(s) respectively; the data for 2 threads has been included in Table 9; lower value indicates better performance.

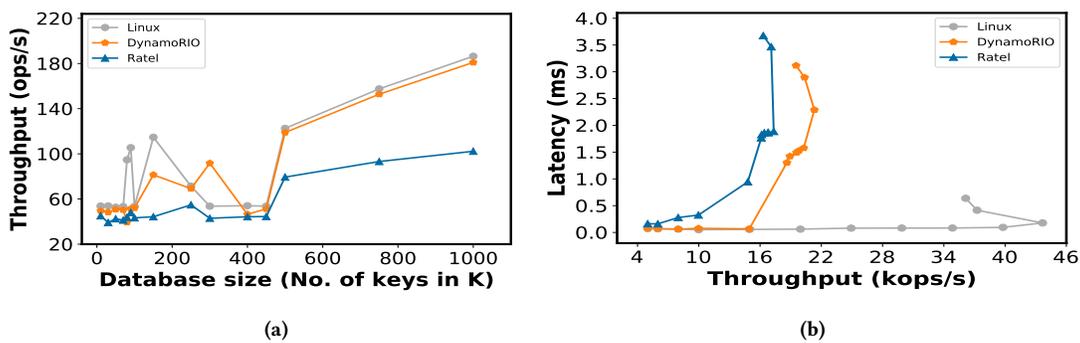

**Figure 10:** RATEL performance for SQLite and Memcached. (a) shows SQLite's average time per operation (micros/op) with increasing database size represented as number of primary keys in thousands (K) across Linux, vanilla DynamoRIO, and RATEL; (b) shows the throughput versus latency of Memcached on Linux, vanilla DynamoRIO, and RATEL.

| Property | Sub-property | Performance | | | | Overhead (in %) | | | |
|---|---|---|---|---|---|---|---|---|---|
| | | Linux | DR | Ratel | Graphene-SGX | DR | Ratel | R-DR | Graphene-SGX |
| Memory Intensive Operations Bandwidth (MB/s) More iteration Less Chunk size | Raw Memory Read | 17915.33 | 24976.73 | 24665.05 | 23210.89 | 39.42 | 37.68 | -1.25 | 29.56 |
| | Raw Memory Write | 9928.48 | 12615.13 | 12580.36 | 12114.76 | 27.06 | 26.71 | -0.28 | 22.02 |
| | Bzero Bandwidth | 62393.29 | 60877.42 | 65072.41 | 47844.32 | -2.43 | 4.29 | 6.89 | -23.32 |
| | Memory copy libc aligned | 41565.35 | 56883.67 | 60377.04 | 63423.44 | 36.85 | 45.26 | 6.14 | 52.59 |
| | Memory copy libc unaligned | 9497.17 | 56270.52 | 61543.81 | 69444.44 | 492.5 | 548.02 | 9.37 | 631.21 |
| | Memory copy unrolled aligned | 9221.96 | 12272.93 | 12351.22 | 12161.04 | 33.08 | 33.93 | 0.64 | 31.87 |
| | Memory copy unrolled unaligned | 9151.18 | 12279.55 | 12295.40 | 10079.86 | 34.19 | 34.36 | 0.13 | 10.15 |
| | Mmapped Read | 706.63 | 423.85 | 190.73 | 3814.69 | -40.02 | -73.01 | -55.00 | 439.84 |
| | File Read | 74.16 | 29.15 | 12.05 | 325.52 | -60.69 | -83.75 | -58.66 | 338.94 |
| Memory Intensive Operations Bandwidth (MB/s) Less iteration More Chunk size | Raw Memory Read | 10708.64 | 13292.24 | 5717.96 | 5310.06 | 24.13 | -46.6 | -56.98 | -50.41 |
| | Raw Memory Write | 9008.42 | 10664.76 | 4563.60 | 3495.86 | 18.39 | -49.34 | -57.21 | -61.19 |
| | Bzero Bandwidth | 21794.54 | 31315.64 | 4166.62 | 4046.79 | 43.69 | -80.88 | -86.69 | -81.43 |
| | Memory copy libc aligned | 12948.37 | 12969.82 | 1570.94 | 1534.59 | 0.17 | -87.87 | -87.89 | -88.15 |
| | Memory copy libc unaligned | 12870.26 | 13141.00 | 1556.58 | 1545.08 | 2.1 | -87.91 | -88.15 | -87.99 |
| | Memory copy unrolled aligned | 6609.18 | 6714.93 | 2054.55 | 2009.46 | 1.6 | -68.91 | -69.40 | -69.6 |
| | Memory copy unrolled unaligned | 6035.19 | 5853.26 | 2081.05 | 1999.58 | -3.01 | -65.52 | -64.45 | -66.87 |
| | Mmapped Read | 4839.57 | 7163.38 | 3299.77 | 1454.30 | 48.02 | -31.82 | -53.94 | -69.95 |
| | File Read | 285.39 | 3724.39 | 769.66 | 134.42 | 1205 | 169.68 | -79.33 | -52.9 |
| File System Latency(us) | Filesystem create | 2.37 | 32.43 | 115.34 | 1272.86 | 1268.35 | 4766.67 | 255.66 | 53607.17 |
| | Filesystem delforward | 0.94 | 18.41 | 33.53 | 1185.10 | 1858.51 | 3467.02 | 82.13 | 125974.47 |
| | Filesystem delrand | 0.92 | 21.17 | 37.28 | 1073.69 | 2201.09 | 3952.17 | 76.10 | 116605.43 |
| | Filesystem delreverse | 0.99 | 18.31 | 33.13 | 1266.38 | 1749.49 | 3246.46 | 80.94 | 127817.17 |
| System Call Latency(us) | getpid | 0.0087 | 0.0065 | 0.0058 | 0.0901 | -25.12 | -33.18 | -10.77 | 938.02 |
| | getrusage | 0.4831 | 0.6401 | 7.4504 | 0.0903 | 32.50 | 1442.21 | 1063.94 | -81.31 |
| | gettimeofday | 0.0276 | 0.0239 | 6.5986 | 6.8500 | -13.28 | 23842.67 | 27509.21 | 24754.86 |
| | sbrk | 0.0076 | 0.0064 | 0.0065 | 0.0102 | -16.23 | -14.92 | 1.56 | 33.51 |
| | sigaction | 0.6269 | 2.2101 | 2.7903 | 0.5904 | 252.56 | 345.11 | 26.25 | -5.82 |
| | write | 0.4927 | 0.5104 | 7.2301 | 0.5303 | 3.59 | 1367.44 | 1316.56 | 7.63 |
| Signal Handler Latency(us) | Installing Signal | 0.48 | 2.24 | 2.79 | 0.60 | 365.75 | 480.11 | 24.55 | 24.76 |
| | Handling Signal | 1.18 | 8.88 | 81.58 | 0.37 | 652.9 | 6816.84 | 818.69 | -68.63 |

**Table 8: Summary of HBenchOS benchmark results for Graphene-SGX along with Linux, DynamoRIO and Ratel.**

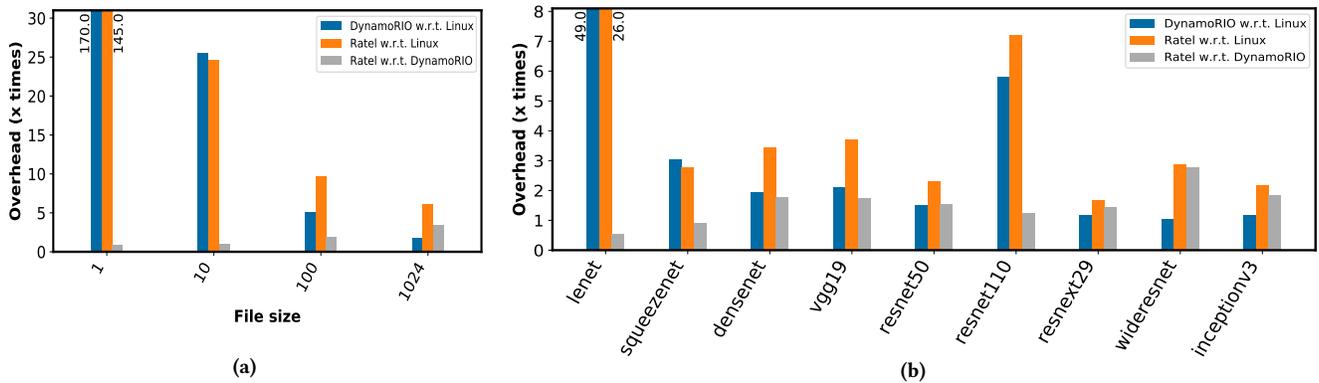

**Figure 11:** Ratel performance for (a) cURL and (b) Privado. Vanilla DynamoRIO execution time w.r.t. Linux, Ratel execution time w.r.t. Linux, and Ratel execution time w.r.t. vanilla DynamoRIO.

Ratel offers instruction-level instrumentation capability. Some performance overheads in Graphene-SGX are expected to be better than Ratel due to the differing design choices. Graphene-SGX does not use spin-locks and tunnels all signal handling through libc as it prioritizes performance over binary compatibility, and has reduced overheads compared to Ratel. On the other hand, Ratel offers better binary compatibility as opposed to Graphene-SGX which provides compatibility with glibc, as shown in Section 6.1.1.

### 6.3.2 Effects of Memory constraints on Ratel.
On our current experimental setup, SGX has a maximum of 128 MB EPC i.e., private physical memory, of which approximately 90 MB is available for user-enclaves. Further, the platform supports at most 64 GB of virtual memory per enclave. These limitations adversely impact Ratel performance.

**Physical Memory Footprint.** We report the physical memory required to execute each application with Ratel (Column 5, Table 9), the smallest being 236 MB for FSCQ benchmark binaries. Ratel uses this memory for the target application as well as for DynamoRIO binaries and the code cache. Since the EPC size is only 90MB, executing enclaves with a physical memory footprint larger than this size (236MB or more in our experiments) causes a high number of page faults (Column 12, Table 9). This is one of the main sources for Ratel performance overheads.

**Virtual Memory Footprint.** We monitor the peak virtual memory usage for Linux and DynamoRIO using ptrace and procmaps. We use sgxtop to monitor the enclave peak virtual memory at run time (Column 8, Table 9 [9]). Further, we monitor the peak virtual memory used by the untrusted host application corresponding to the enclave (Column9, Table 9). On average, DynamoRIO incurs a high memory overhead of 205× compared to Linux. However, Ratel only imposes 24× overhead compared to Linux. There are two reasons why Ratel incurs significantly lower virtual memory usage compared to DynamoRIO.

First, DynamoRIO reserves 2 GB of heap memory region as a scalability improvement for Linux x64 [5]. Our analysis of DynamoRIO with several binaries shows that this region is rarely used. SGX has a limited EPC and requires pre-specifying maximum heap size. We disable this reservation logic in Ratel to reduce its virtual memory usage and subsequent page faults for loading physical pages. Further, a low memory footprint speeds up the enclave creation and attestation because SGX has to initialize and measure a smaller memory region. Thus, Ratel virtual memory peak is always smaller than DynamoRIO by at least a margin of 2 GB (see Column 7 and 8 in Table 9).

Second, DynamoRIO executes directly on Linux. Thus, it can demand arbitrarily large physical memory (as long as it is available on the RAM). It uses the default Linux memory manager to optimize memory allocations. On the other hand, Ratel has to pre-specify the maximum physical footprint before execution and it uses a modified SGX SDK memory manager for its own heap (see Section 4.2). We initialize just enough enclave memory such that Ratel can execute target applications and not any more than that. Such conservative allocation allows Ratel to quickly create and launch enclaves. This explains why Ratel has lower virtual memory peaks (typically between 256 MB to 2 GB) compared to DynamoRIO (1–2 GB, after accounting for 2 GB for the above optimizations). To empirically verify our hypothesis, we perform a controlled experiment. For each binary from SPEC 2006, we run Ratel with increasing size of maximum heap memory (ranging from 256 MB to 4 GB) and measure the virtual memory peak. We report that the virtual memory peak continues to increase with increasing maximum heap size and then it plateaus at a certain point. The plateau point of each binary matches the corresponding DynamoRIO peak. This confirms our claim that Ratel has a smaller virtual memory peak because we limit the maximum heap size in our configuration. These two phenomena explain why Ratel has a much smaller virtual memory footprint compared to DynamoRIO.

Note that Ratel additionally incurs virtual memory overheads in the untrusted host application (Column 9, Table 9). The two copy model used in Ratel design accounts for a high virtual memory peak in the untrusted part of the process.

**Code Cache Size.** Ratel and DynamoRIO allow users to configure the maximum code cache size via a configuration file before launching the enclave. The cache is used to store basic blocks and traces. At run time, the DBT engine is allowed to use a cache up to this size. Often, the peak cache size is smaller than the maximum because the basic blocks and traces may fit in a smaller memory for a given application. We execute applications with different code cache sizes for both DynamoRIO and Ratel. Our tests start with a maximum cache size of 4 KB and we double the size up to 64 MB. For each run, we measure the peak basic block cache and trace cache size in DynamoRIO and Ratel.

Once the maximum cache size is large enough for the application, both of them execute successfully. Further, the peak size stays constant even if we keep increasing the maximum size. In the case of memcached executing YCSB workload A, DynamoRIO peaks at 231.24 KB basic block cache and 92.78 KB for the trace cache. Ratel peaks at 201.54 KB and 32.84 KB respectively. Increasing the cache size beyond the peak value does not improve the performance of DynamoRIO or Ratel. Specifying a large cache size for Ratel results in a larger enclave physical memory. Ratel takes more time to initialize and create the enclave, it also incurs more frequent page faults. Thus, beyond the peak size, these two factors slow-down the application with an increase in the cache size. When we reduce the cache size below the peak value, DynamoRIO suffers an order of magnitude slowdown. This is a well-known and expected behavior [25]. Relatively, Ratel does not suffer such a drastic slowdown, partly because a smaller cache size results in fewer page faults. However, if the specified code cache size is very small compared to the peak value, Ratel fails to execute a given application.

In summary, the limited EPC size in SGX v1 not only results in high execution overhead but also invalidates expected performance gains via a large code cache.

## 6.4 Compatibility with Built-in Profilers

Ratel primarily achieves binary compatibility by leveraging the complete interposition offered by DynamoRIO. Additionally, this instruction-level interposition allows Ratel to monitor various in-enclave behavior (e.g., events, instructions, control-flows, etc.) out-of-the-box. Specifically, DynamoRIO provides 26 built-in profilers for dynamically tracing, analyzing, and fine-tuning the target application. Table 10 summarizes the names and the profiling services they offer. When we run vanilla DynamoRIO on our experiment platform, 25/26 profilers work stably. One profiler, -prof_pcs, is unstable and causes time-outs. This is a well-documented issue with DynamoRIO [10, 11]. Of all the 25 profilers that work with DynamoRIO, Ratel retains support for all of them. We experimentally demonstrate that Ratel maintains compatibility with built-in profilers. We randomly choose 4 application binaries from each of the 6 categories listed in Table 9. We run a total of 24 applications that exhibit diverse execution behavior with all of the 25 profilers. We report that all of 25 profilers worked with all our applications.

| Suite Name | Benchmark /Application Name | Compile Stats | | | Runtime Stats | | | | | | | | Time (sec) | | | Overhead (in %) | | |
|---|---|---|---|---|---|---|---|---|---|---|---|---|---|---|---|---|---|---|
| | | LOC | Binary Size | Mem Size | Linux VmP. | DR VmP. | Trust. VmP. | Untru. VmP. | Out Calls | Sys Calls | Page Faults | Ctx Swt | Linux | DR | Ratel | DR | Ratel | R-DR |
| SPEC CINT2006 | astar | 4280 | 56 KB | 237.68 | 21.48 | 3101.62 | 257.94 | 101.34 | 26561 | 618555 | 271544 | 181 | 7.25 | 8.77 | 10.71 | 20.97 | 47.72 | 22.01 |
| | bzip2 | 5734 | 73 KB | 519.05 | 208.98 | 3289.13 | 1024 | 248.13 | 26048 | 618115 | 443869 | 238 | 19.21 | 21.74 | 34.49 | 13.17 | 79.54 | 58.99 |
| | gobmk | 157650 | 4.4 MB | 244.84 | 35.70 | 3115.85 | 258.07 | 107.43 | 26594 | 618629 | 272957 | 150 | 0.82 | 1.73 | 4.37 | 110.98 | 432.93 | 152.60 |
| | hmmer | 20680 | 331 KB | 238.25 | 7.93 | 3088.11 | 258.62 | 101.66 | 26144 | 629203 | 271106 | 139 | 0.19 | 0.98 | 0.85 | 415.79 | 347.37 | -13.27 |
| | sjeng | 10549 | 162 KB | 355.39 | 178.86 | 3259.90 | 512 | 217.64 | 26606 | 618638 | 2049404 | 382 | 2.98 | 5.08 | 4.57 | 70.47 | 53.36 | -10.04 |
| | libquantum | 2611 | 51 KB | 237.29 | 7.32 | 3087.46 | 257.1 | 101.33 | 25969 | 618004 | 271071 | 150 | 0.03 | 0.50 | 0.47 | 1566.67 | 1466.67 | -6.75 |
| | h264ref | 36097 | 602 KB | 240.59 | 34.64 | 3114.79 | 259.16 | 102.18 | 27033 | 619033 | 272100 | 284 | 8.87 | 18.25 | 34.83 | 105.75 | 292.67 | 90.16 |
| | omnetpp | 26652 | 871 KB | 239.82 | 20.67 | 3100.81 | 258.42 | 102 | 26961 | 618990 | 271813 | 151 | 0.26 | 2.47 | 2.72 | 850.00 | 946.15 | 10.12 |
| | Xalan | 267376 | 6.3 MB | 248.73 | 23.53 | 3103.68 | 261.93 | 105.84 | 28121 | 620185 | 273953 | 198 | 0.05 | 5.53 | 4.05 | 10960.00 | 8000.00 | -26.76 |
| | gcc | 385783 | 3.8 MB | 251.16 | 20.17 | 3100.28 | 265.15 | 105.44 | 25758 | 656241 | 56201 | 454 | 5.38 | 12.85 | 6.30 | 138.85 | 17.10 | -51.16 |
| | gromac | 87921 | 1.1 MB | 239.91 | 24.17 | 3104.32 | 259.62 | 102.28 | 26783 | 654600 | 55019 | 633 | 0.48 | 2.85 | 4.79 | 493.75 | 897.92 | 68.07 |
| SPEC CFP2006 | leslie3d | 2983 | 177 KB | 238.86 | 28.49 | 3108.63 | 258.54 | 101.39 | 26831 | 618865 | 271723 | 204 | 5.25 | 18.80 | 21.63 | 258.10 | 312.00 | 14.89 |
| | milc | 9580 | 150 KB | 238.41 | 16.04 | 3096.18 | 256 | 101.45 | 32551 | 624587 | 271506 | 192 | 7.27 | 13.05 | 22.41 | 79.50 | 208.25 | 70.99 |
| | namd | 3892 | 330 KB | 238.46 | 58.14 | 3138.29 | 256 | 101.6 | 28550 | 620582 | 271665 | 173 | 8.13 | 18.86 | 19.41 | 131.98 | 138.75 | 2.65 |
| | cactusADM | 60235 | 819 KB | 596.88 | 415.05 | 3495.19 | 1027.35 | 454.19 | 27619 | 619634 | 370217 | 190 | 1.13 | 5.34 | 7.78 | 372.57 | 588.50 | 45.69 |
| | calculix | 105123 | 1.8 MB | 395.25 | 169.34 | 3249.48 | 512 | 208.48 | 27319 | 629243 | 313313 | 174 | 0.03 | 3.06 | 3.90 | 10000.00 | 12900.00 | 27.45 |
| | dealII | 94458 | 4.3 MB | 277.37 | 97.08 | 3177.76 | 515.73 | 138.61 | 26858 | 618872 | 273471 | 240 | 10.29 | 24.68 | 24.73 | 139.84 | 140.33 | 0.00 |
| | GemsFDTD | 4883 | 440 KB | 1021.82 | 841.85 | 3924.16 | 2048 | 883.48 | 25207 | 617226 | 366889 | 177 | 1.24 | 5.44 | 2.08 | 338.71 | 67.74 | -61.76 |
| | povray | 78684 | 1.2 MB | 242.70 | 16.54 | 3096.69 | 262.25 | 102.47 | 29082 | 621108 | 272267 | 166 | 0.35 | 4.42 | 4.72 | 1162.86 | 1248.57 | 6.79 |
| | soplex | 28282 | 507 KB | 239.68 | 16.90 | 3097.04 | 256 | 101.71 | 26880 | 618909 | 271861 | 164 | 0.01 | 2.07 | 2.08 | 20600.00 | 20700.00 | 0.48 |
| | specrand (998) | 54 | 8.7 KB | 236.76 | 4.25 | 3084.39 | 256.01 | 101.3 | 25863 | 617897 | 270924 | 150 | 0.23 | 0.35 | 0.27 | 52.17 | 17.39 | -22.41 |
| | specrand (999) | 54 | 8.7 KB | 236.76 | 4.24 | 3084.39 | 256 | 101.3 | 25863 | 617897 | 270990 | 164 | 0.21 | 0.34 | 0.33 | 61.90 | 57.14 | -2.94 |
| | tonto | 107228 | 4.6 MB | 248.28 | 20.54 | 3100.68 | 264.75 | 105.57 | 30562 | 622574 | 273669 | 186 | 0.43 | 6.89 | 6.65 | 1502.33 | 1446.51 | -3.48 |
| | zeusmp | 19030 | 280 KB | 1357.84 | 1132.98 | 4213.15 | 2051.73 | 1219.49 | 27163 | 619021 | 1755130 | 434 | 7.55 | 20.65 | 52.03 | 173.51 | 589.14 | 152.43 |
| IOZONE | read/reread | 26545 | 1.1 MB | 241.64 | 48.23 | 3128.37 | 257.86 | 105.12 | 27254 | 622791 | 23785 | 1215 | 0.06 | 0.88 | 0.89 | 1366.67 | 1383.33 | 1.14 |
| | random r./w. | 26545 | 1.1 MB | 241.64 | 48.23 | 3128.37 | 257.86 | 105.12 | 27376 | 622913 | 23744 | 844 | 0.07 | 0.88 | 1.09 | 1157.14 | 1457.14 | 23.86 |
| | backward read | 26545 | 1.1 MB | 241.64 | 48.23 | 3128.37 | 257.86 | 105.12 | 27431 | 622968 | 23854 | 1159 | 0.07 | 0.84 | 1.38 | 1100.00 | 1871.43 | 64.29 |
| | fwrite/frewrite | 26545 | 1.1 MB | 241.64 | 48.36 | 3128.50 | 257.86 | 105.12 | 27212 | 622750 | 24317 | 581 | 0.07 | 0.87 | 0.86 | 1142.86 | 1128.57 | -1.15 |
| | fread/freread | 26545 | 1.1 MB | 241.64 | 48.36 | 3128.50 | 257.86 | 105.12 | 27223 | 622760 | 23742 | 374 | 0.06 | 0.86 | 0.86 | 1333.33 | 983.33 | -24.42 |
| FSCQ | fscq large file | 383 | 25 KB | 236.82 | 4.26 | 3084.40 | 256.27 | 101.3 | 25889 | 1165892 | 270914 | 168 | 0.12 | 0.47 | 3.41 | 291.67 | 2741.67 | 625.53 |
| | fscq small file | 161 | 19 KB | 236.82 | 4.29 | 3084.43 | 256.19 | 101.34 | 26352 | 929795 | 270959 | 181 | 0.01 | 0.34 | 0.17 | 3300.00 | 1600.00 | -50.00 |
| | fscq write file | 74 | 18 KB | 236.82 | 4.25 | 3084.39 | 256.04 | 101.3 | 262015 | 930226 | 270867 | 143 | 0.01 | 0.31 | 0.13 | 3000.00 | 1200.00 | -58.06 |
| | multicreatewrite | 20 | 11 KB | 236.81 | 4.24 | 3084.39 | 257.5 | 101.3 | 65721 | 969595 | 270969 | 248 | 0.11 | 0.38 | 0.83 | 245.45 | 654.55 | 120.74 |
| | multiopen | 14 | 9.8 KB | 236.81 | 4.24 | 3084.39 | 257.5 | 101.3 | 225719 | 1129593 | 270842 | 452 | 0.16 | 0.57 | 2.44 | 256.25 | 1425.00 | 328.07 |
| | multicreate | 18 | 9.9 KB | 236.81 | 4.24 | 3084.39 | 257.5 | 101.3 | 55720 | 949625 | 270866 | 212 | 0.07 | 0.33 | 0.67 | 342.86 | 857.14 | 116.13 |
| | multiwrite | 16 | 9.9 KB | 236.81 | 4.24 | 3084.39 | 257.5 | 101.3 | 35720 | 939594 | 270864 | 152 | 0.01 | 0.23 | 0.30 | 2200.00 | 2900.00 | 28.21 |
| | multicreatemany | 19 | 11 KB | 236.81 | 4.24 | 3084.39 | 257.5 | 101.3 | 45729 | 959605 | 271034 | 198 | 0.07 | 0.36 | 0.77 | 414.29 | 1000.00 | 115.08 |
| | multiread | 17 | 9.9 KB | 236.81 | 4.24 | 3084.39 | 257.5 | 101.3 | 325721 | 1229595 | 270901 | 589 | 0.21 | 0.62 | 3.70 | 195.24 | 1661.90 | 494.86 |
| PARSEC-SPLASH2 | water_nsquare | 2885 | 46 KB | 239.27 | 18.07 | 3098.21 | 258.53 | 129.22 | 27109 | 622992 | 25093 | 199 | 0.05 | 0.94 | 0.88 | 1780.00 | 1660.00 | -6.08 |
| | water_spatial | 3652 | 46 KB | 456.63 | 145.71 | 3225.85 | 514.88 | 256.86 | 27171 | 622991 | 61992 | 164 | 0.04 | 1.05 | 2.31 | 2525.00 | 5675.00 | 120.00 |
| | barnes | 4942 | 46 KB | 247.13 | 67.42 | 3147.56 | 257.68 | 178.57 | 26801 | 622748 | 28191 | 287 | 0.19 | 0.85 | 1.02 | 347.37 | 436.84 | 20.14 |
| | fmm | 7611 | 64 KB | 455.38 | 146.11 | 3226.25 | 513.14 | 257.26 | 27218 | 622909 | 62025 | 58 | 0.01 | 0.97 | 2.18 | 9600.00 | 21700.00 | 125.44 |
| | raytrace | 200091 | 92 KB | 455.63 | 186.00 | 3266.15 | 512.05 | 297.15 | 27291 | 623175 | 65323 | 194 | 0.05 | 1.23 | 2.59 | 2360.00 | 5080.00 | 110.57 |
| | radiosity | 21586 | 230 KB | 455.63 | 63.15 | 3143.30 | 512.01 | 174.3 | 27609 | 623118 | 26162 | 139 | 1.92 | 4.30 | 5.82 | 123.96 | 203.13 | 35.35 |
| | ocean_cp | 10519 | 81 KB | 238.63 | 31.76 | 3111.91 | 258.31 | 142.91 | 27234 | 622990 | 25480 | 86 | 0.05 | 1.19 | 1.05 | 2280.00 | 2000.00 | -11.76 |
| | ocean_ncp | 6275 | 65 KB | 238.25 | 40.36 | 3120.51 | 257.11 | 151.51 | 27052 | 622948 | 25362 | 91 | 0.05 | 1.03 | 1.08 | 1960.00 | 2060.00 | 4.85 |
| | volrend | 27152 | 271 KB | 238.00 | 146.02 | 3226.17 | 256.01 | 129.17 | 27309 | 623167 | 25082 | 178 | 0.01 | 0.75 | 0.88 | 7400.00 | 8700.00 | 17.02 |
| Applications | SQLite(10K keys) | 140420 | 1.3 MB | 241.24 | 24.58 | 3104.72 | 256 | 101.39 | 400548 | 1181195 | 272323 | 1261 | 5.05 | 5.82 | 6.94 | 15.25 | 37.43 | 19.24 |
| | cURL (10 MB) | 22064 | 30 KB | 266.07 | 76.77 | 3156.91 | 512.95 | 127.22 | 35897 | 940802 | 272552 | 1031 | 0.07 | 1.78 | 1.17 | 2442.86 | 1571.43 | -34.27 |
| | Memcach(100K) | 44921 | 795 KB | 1589.42 | 595.55 | 3547.69 | 2048 | 1408.99 | 1021241 | 1118961 | 540649 | 104765 | 5.28 | 5.99 | 9.46 | 13.45 | 79.17 | 57.93 |
| | densenetapp | 12551 | 32 MB | 752.04 | 575.70 | 3655.41 | 1028.5 | 614.41 | 27826 | 616749 | 123894 | 354 | 3.74 | 7.25 | 12.90 | 93.85 | 244.92 | 77.93 |
| | lenetapp | 230 | 313 KB | 237.29 | 8.12 | 3088.26 | 256.38 | 101.59 | 26029 | 614611 | 21166 | 362 | 0.01 | 0.49 | 0.26 | 4800.00 | 2500.00 | -47.05 |
| | resnet110app | 9528 | 110 MB | 270.43 | 94.42 | 3175.39 | 512 | 134.38 | 27238 | 696291 | 23716 | 200 | 0.34 | 1.98 | 2.45 | 482.35 | 620.59 | 23.74 |
| | resnet50app | 2826 | 98 MB | 605.50 | 430.67 | 3511.61 | 1025.7 | 470.62 | 26591 | 616291 | 136139 | 274 | 5.09 | 7.64 | 11.81 | 50.10 | 132.02 | 54.45 |
| | resnext29app | 1753 | 132 MB | 575.01 | 400.85 | 3481.00 | 1025.08 | 439.99 | 26410 | 616728 | 187284 | 411 | 9.76 | 11.38 | 16.33 | 16.60 | 67.32 | 42.98 |
| | squeezenetapp | 914 | 4.8 MB | 242.15 | 59.41 | 3135.59 | 258.15 | 106.07 | 26258 | 616290 | 23001 | 252 | 0.40 | 1.22 | 1.11 | 205.00 | 177.50 | -9.02 |
| | vgg19app | 990 | 77 MB | 345.78 | 171.65 | 3252.65 | 514.15 | 211.46 | 26192 | 630872 | 96647 | 402 | 0.66 | 1.40 | 2.44 | 112.12 | 269.70 | 74.29 |
| | widgeresnetapp | 1495 | 140 MB | 564.14 | 390.19 | 3470.72 | 1025.43 | 429.71 | 26352 | 631004 | 172712 | 303 | 19.25 | 20.02 | 55.38 | 4.00 | 187.69 | 177.00 |
| | inception3 | 4875 | 92 MB | 656.53 | 481.82 | 3561.96 | 1024 | 520.96 | 26862 | 1088344 | 250880 | 355 | 11.39 | 13.25 | 24.63 | 16.33 | 116.24 | 84.96 |

**Table 9:** Ratel statistics for benchmarks and real-world applications. Columns $3 - 4$: total application LoC and binary size. Columns 5: maximum physical memory size (in MB) required to execute each application with Ratel. Columns $6 - 7$: peak virtual memory usage (in MB) on Linux and DynamoRIO. Columns $8 - 9$: trusted and untrusted peak virtual memory usage (in MB) on Ratel. Columns $10 - 13$: total OCALLs, system calls, page faults, and context switches recorded in one run. Columns $14 - 16$: execution time on Linux, vanilla DynamoRIO, and Ratel. Column $17 - 19$: execution overhead of DynamoRIO w.r.t. Linux, Ratel w.r.t. Linux, and Ratel w.r.t. DynamoRIO. Ratel performs better than DynamoRIO in some cases (denoted by negative overheads)—SGX loads the binaries during enclave attestation and we do not include the enclave creation time in Ratel execution time.

| Built-in Profiler | Support | Description |
|---|---|---|
| -opt_memory | ✓ | Reduce memory usage, but potentially at the cost of performance. |
| -prof_pcs | ✗ | A simple sampler to periodically interrupt DBT engine and query which part of DBT engine was running. |
| -stack_size \<number> | ✓ | Increase the size of DBT engine's per-thread stack. |
| -signal_stack_size \<number> | ✓ | Specify the size of signal handling stack. |
| -thread_private | ✓ | Request code caches that are private (shared across threads by default) to each thread. |
| -disable_traces | ✓ | Disable trace building (e.g., basic block cache and trace cache), which can have a negative performance impact. |
| -enable_full_api | ✓ | Default internal options balance performance with API usability. |
| -max_bb_instrs | ✓ | Stop building a basic block if it hits this application instruction count limit. |
| -max_trace_bbs | ✓ | Build a trace with less than this number of constituent basic block. |
| -synch_at_exit | ✓ | In debug builds, synchronize with all remaining threads at process exit time. |
| -syntax_intel | ✓ | Output all disassembly using Intel syntax rather than the default AT&T-style syntax. |
| -tracedump_text | ✓ | A text dump option to output all traces that were created to the log file traces-shared.0.TID.html. |
| -tracedump_binary | ✓ | A binary dump option to output all traces that were created to the log file traces-shared.0.TID.html. |
| -tracedump_origins | ✓ | Dump only a text list of the constituent basic block tags of each trace to the trace log file. |
| -reachable_heap | ✓ | Guarantee all of the heap memory is reachable from the code cache, at the risk of running out of memory. |
| -multi_thread_exit | ✓ | Avoid synchronizing with all remaining threads at process exit time. |
| -cache_bb_max | ✓ | Set maximum basic block code cache sizes. |
| -cache_trace_max | ✓ | Set maximum trace code cache sizes. |
| -msgbox_mask 0xN | ✓ | Control whether the system waits for a key press, when presenting information. |
| -stderr_mask 0xN | ✓ | Control the output to standard error. |
| -pause_on_error | ✓ | Suspend the process so that a debugger can be attached when encountering an assert or crash. |
| -debug | ✓ | Use the DBT engine debug library for debugging. |
| -loglevel N | ✓ | Print out a log of DBT engine's actions. The greater the value of N, the more information the system prints. |
| -logmask 0xN | ✓ | Select which DBT engine modules print out logging information, at the -loglevel level. |
| -ignore_assert_list '*' | ✓ | Ignore all DBT engine asserts of the form "\<file>:1234". |
| -logdir \<path> | ✓ | Specify the directory to use for log files. |

Table 10: Names and description of built-in Profilers in DynamoRIO that are available directly in RATEL. ✓ indicates that the profiler is supported out-of-the-box in RATEL and we have tested it successfully. ✗ indicates that the profiler is not supported in RATEL because it crashes in vanilla DynamoRIO [10, 11]

## 7 RELATED WORK

Several prior works have targeted SGX compatibility. There are two main ways that prior work has overcome these challenges. The first approach is to fix the application interface. The target application is either re-compiled or is relinked to use such interfaces. The approach that enables the best compatibility exposes specific Libc (glibc or musl libc) versions as interfaces. This allows them to adapt to SGX restrictions at a layer below the application. Container or library OS solutions use this to execute re-compiled/re-linked code inside the enclave as done in Haven [19], Scone [15], Graphene-SGX [28], Ryoan [46], SGX-LKL [65], and Occlum [69]. Another line of work is compiler-based solutions. They require applications to modify source code to use language-level interface [38, 59, 72, 84].

Both style of approaches can have better performance than RATEL, but require recompiling or relinking applications. For example, library OSes like Graphene-SGX and containerization engines like Scone expose a particular glibc and musl version that applications are asked to link with. New library versions and interfaces can be ported incrementally, but this creates a dependence on the underlying platform interface provider, and incurs a porting effort for each library version. Applications that use inline assembly or runtime code generation also become incompatible as they make direct access to system calls, without using the API. RATEL approach of handle R1-R5 comprehensively offers complete interposition, without any assumptions about specific interfaces beyond that implied by binary compatibility.

**Security Considerations.** As in RATEL, other approaches to SGX compatibility eventually have to use OCALLs, ECALLs, and syscalls to exchange information between the enclave and the untrusted software. This interface is known to be vulnerable [29, 79]. Several shielding systems for file [26, 74] and network IO [17], provide specific mechanisms to safeguard the OS interface against these attacks. For security, defense techniques offer compiler-based tools for enclave code for memory safety [51], ASLR [68], preventing controlled-channel leakage [70], data location randomization [21], secure page fault handlers [63], and branch information leakage [44].

**Performance.** Several other works build optimizations by modifying existing enclave-compliant library OSes. One such example is Hotcalls [86], Eleos [62] which add exit-less calls to reduce the overheads of OCALLs. These optimizations are also available in the default Intel SGX SDK now.

**Language Run-times.** Recent body of work has also shown how executing either entire [83] or partial [31] language runtimes inside an enclave can help to port existing code written in interpreted languages such as Python [57, 66], Java [31], web-assembly [42], Go [41], and JavaScript [43].

**Programming TEE Applications.** Intel provides a C/C++ SGX software stack which includes a SDK and OS drivers for simulation and PSW for running local enclaves. There are other SDKs developed in in memory safe languages such as Rust [38, 59, 84]. Frameworks such as Asylo [16], OpenEnclave [61], and MesaTEE [58] expose a high-level front-end for writing native TEE applications using a common interface. They support several back-end TEEs including Intel SGX and ARM TrustZone. Many of the challenges faced by RATEL are common to these frameworks.

**Future TEEs.** New enclave TEE designs have been proposed [27, 34, 36, 52, 73]. Micro-architectural side channels [20] and new oblivious execution capabilities [34, 54] are significant concerns in these designs. Closest to our underlying TEE is the recent Intel SGX v2 [12, 55, 87]. SGX v2 enables dynamically memory and thread management inside the enclave, thus addressing $R2$ to some extent. The other restrictions are largely not addressed in SGX v2, and therefore, RATEL design largely applies to it as well.

**SGX v2 & Larger EPC.** SGX v2 offers advantages which can improve the security and performance of RATEL. In SGX v1, the enclave pages can never be dynamically added, deleted, or have modified permissions after enclave initialization. v2 removes these restrictions with new SGX instruction support. Specifically, the readable+writable code cache in RATEL can be protected from by dynamic permission changes as done in the vanilla DynamoRIO). However, adding or removing the execute permission to enclave pages is disallowed even on SGX v2. Newer machines that support SGX will have larger available EPC size. Thus, such a larger code cache capacity can speedup RATEL enclave creation and application execution. Thus, the page swapping frequency will reduce especially for large legacy binaries. This may reduce the runtime overhead of RATEL. However, all the performance bottlenecks due to R1, R3, R4, R5 will manifest even on SGX v2.

## 8 CONCLUSION

We present the design of RATEL, which enables dynamic binary translation inside SGX enclaves. This offers the ability to interpose on all instruction executed in an enclave, which serves as a foundation for implementing other security monitors to safeguard enclaves from bugs and from the untrusted OS. RATEL also provides the first evidence that binary compatibility with existing Linux software on SGX is feasible. We empirically report on an extensive evaluation with over 200 common Linux applications and multiple scripting language runtimes. Our observations about the restrictive design choices made in SGX may be of independent interest to designers of next-generation enclave systems.

## ACKNOWLEDGMENTS

We thank David Kohlbrenner, Zhenkai Liang, and Roland Yap for their feedback on improving earlier drafts of the paper. We thank Shipra Shinde for help on formatting the figures in this paper. This research was partially supported by a grant from the National Research Foundation, Prime Ministers Office, Singapore under its National Cybersecurity R&D Program (TSUNAMi project, No. NRF2014NCR-NCR001-21) and administered by the National Cybersecurity R&D Directorate. This material is in part based upon work supported by the National Science Foundation under Grant No. DARPA N66001-15-C-4066 and Center for Long-Term Cybersecurity. Any opinions, findings, and conclusions or recommendations expressed in this material are those of the authors and do not necessarily reflect the views of the National Science Foundation.

## A  PERFORMANCE BREAKDOWN

### A.1  Detailed Breakdown for Micro-benchmarks

We measure the performance on diverse workloads, which helps us to explain the costs associated with executing with RATEL.

**System Stress Workloads.** We use HBenchOS [22]—a benchmark to measure the performance of primitive functionality provided by an OS and hardware platform. In Table 8 we show the cost of each system-level operation such as system calls, memory operations, context switches, and signal handling. Memory-intensive operation latencies vary with benchmark setting: (a) when the operations are done with more iterations (in millions) and less memory chunk size (4 KB) the performance is comparable; (b) when the operations are done with less iterations (1 K) and more memory chunk size (4 MB) RATEL incurs bandwidth loss ranging from −169.68% to 87.91% and 53.94% to 88.15% over Linux and DynamoRIO, respectively. This happens because when the chunk size is large, we need to allocate and de-allocate memory inside enclave for every iteration as well as copy large amounts of data.

These file operation latencies match with latencies we observed in our I/O intensive workloads (Figure 8). Specifically, the write operation incurs large overhead. Hence, the `create` workload incurs 4766.67% and 255.66% overhead over Linux and DynamoRIO because the benchmark creates a file and then writes predefined sized data to it. Costs of system calls that are executed as OCALLs vary depending on return value and type of the system call. For example, system calls such as `getpid`, `sbrk`, `sigaction` that return integer values are much faster. Syscalls such as `getrusage`, `gettimeofday` returns structures or nested structures. Thus, copying these structures back and forth to/from enclaves causes much of the performance slowdown. RATEL has a custom mechanism for registering and handling signal (Section 4.5); it introduces a latency of 480.11% and 6816.84% with respect to Linux as well as 24.55% and 818.69% with respect to DynamoRIO respectively. Registering signals is cheaper because it does not cause a context switch as in the case of handling the signal. Further, after accounting for the OCALL costs, our custom forwarding mechanism does not introduce any significant slowdown.

**CPU-bound Workloads.** RATEL incurs 217.80% and 34.91% overhead averaged over 24 applications from SPEC 2006 [75] with respect to Linux and DynamoRIO, respectively. Table 9 shows the individual overheads for each application with respect to all baselines. From Table 9 we observe that applications that incur higher number of page faults and OCALLs suffer larger performance slowdowns. Thus, similar to other SGX frameworks, the costs of enclave context switches and limited EPC size are the main bottlenecks in RATEL.

**IO-bound Workloads.** RATEL performs OCALLs for file I/O by copying read and write buffers to and from enclave. We measure the per-API latencies using FSCQ suite for file operations [30]. Table 9 shows the costs of each file operation and file access patterns respectively. Apart from the cost of the OCALL, writes are more expensive compared to reads in general; the multiple copy operations in RATEL amplify the performance gap between them. Next we use IOZone [60], a commonly used benchmark to measure the file I/O latencies. Figure 8 shows the bandwidth over varied file sizes between 16 MB to 1024 MB and record sizes between 4 KB to 4096 KB for common patterns. The trend of writes being more expensive holds for IOZone too. RATEL incurs an average slowdown of 87.5% and 66.2% over all operations, record sizes, and file sizes with respect to Linux and DynamoRIO, respectively.

**Multi-threaded Workloads.** We use the standard Parsec-SPLASH 2 [64] benchmark suite. It comprises a variety of high performance computing (HPC) and graphics applications. We use it to benchmark RATEL overheads for multi-thread applications. Since some of the programs in Parsec-SPLASH2 mandate the thread count to be power to 2 (e.g., ocean_ncp), we fixed the maximum number of threads in our experiment to 16. RATEL changes the existing SGX design to handle thread creation and synchronization primitives, as described in Section 4.3 and 4.4. We measure the effect of this specific change on the application execution by configuring the enclave to use varying number of threads between 1-16. The data for 2 threads is shown in Table 9.

Figure 9 shows a performance overhead of 10156.52% and 92.55% compared to Linux and DynamoRIO, on average, across all benchmarks and thread configurations. Particularly, DynamoRIO imposes an average overhead of 5010.07% over Linux with the same setting. For single-threaded execution, on average, RATEL causes an overhead of 2050.92% and 3.17% with respect to Linux and DynamoRIO, respectively, while they increase to 21238.0% and 159.78% in 16-thread execution. We measure the breakdown of costs and observe that, on average: (a) creating each thread contributes to a fixed cost of 57 ms; (b) shared access to variables becomes expensive by a factor of $1 - 7$ times compared to the elapsed time of `futex` synchronization with increase in number of threads. This is expected because synchronization is cheaper in Linux and DynamoRIO execution, in which they use unsafe `futex` primitives exposed by the kernel. On the other hand, RATEL uses expensive spinlock mechanism exposed by SGX hardware for security. Particularly, some of the individual benchmarks, such as `water_spatial`, `fmm` and `raytrace` that involve lots of lock contention events and have extremely high frequency of spinlock calls (e.g., the spinning counts of about $423,000$ ms in RATEL while the `futex` calls of about 500 ms in DynamoRIO for the raytrace with 8 threads). Thus, they incur large overheads in synchronization.

## A.2 Real-world Case-studies

We work with 4 representative real-world applications: a database server (SQLite), a command-line utility (cURL), a machine learning inference-as-a-service framework (Privado), and a key-value store (memcached). These applications have been used in prior work [72].

**SQLite** is a popular database [76]. We select it as a case-study because of its memory-intensive workload. We configure it as a single-threaded instance. We use a database benchmark workload [53] and measured the throughput (ops/sec) for each database operation with varying sizes of the database (total number of entries). Table 9 shows the detailed breakdown of the runtime statistics for a database with $10,000$ entries. Figure 10a shows the average throughput over all operations. With RATEL, we observe a throughput loss of $36.88\%$ and $28.71\%$ on average over all database sizes compared to Linux and DynamoRIO. The throughput loss increases with increase in the database size. The drop is noticeable at $500K$ where the database size crosses the maximum enclave size threshold and results in significant number of page faults. This result matches with observations from other SGX frameworks that report SQLite performance [15].

**cURL** is a widely used command line utility to download data from URLs [35]. It is network intensive. We test it with RATEL via the standard library test suite. Table 9 shows detailed breakdown of the execution time on RATEL. We measure the cost executing cURL with RATEL for downloading various sizes of files from an Apache (2.4.41) server on the local network. Figure 11a shows the throughput for various baselines and file sizes. On average, RATEL causes a loss of $604.11\%$ and $142.11\%$ throughput as compared to Linux and DynamoRIO. For all baselines, small files (below 100 MB) have smaller download time; larger file sizes naturally take longer time. This can be explained by the direct copying of packets to non-enclave memory, which does not add any memory pressure on the enclave. The only remaining bottleneck in the cost of dispatching `OCALLs` which increase linearly with the requested file size.

**Privado** is a machine learning framework that provides secure inference-as-a-service [44]. It comprises of several state of the art models available as binaries that can execute on an input image to predict its class. The binaries are CPU intensive and have sizes ranging from 313 KB to 140 MB (see Table 9). We execute models from Privado on all the images from the corresponding image dataset (CIFAR or ImageNet) and measure inference time. Figure 11b shows the performance of baselines and RATEL for 9 models in increasing order of binary size. We observe that RATEL performance degrades with increase in binary size. This is expected because the limited enclave physical memory leads to page faults. Hence, largest model (140 MB) exhibit highest inference time and smallest model (313 KB) exhibit lowest inference time. Thus, RATEL and enclaves in general can add significant overheads, even for CPU intensive server workloads, if they exceed the working set size of 90 MB.

**Memcached** is an in-memory key-value cache. We evaluate it with YCSB's all four popular workloads A (50% read and 50% update), B (95% read and 5% update), C (100% read) and D (95% read and 5% insert). We run it with 4 default worker threads running in Linux, DynamoRIO and RATEL settings. We vary the YCSB client threads with *Load* and *Run* operations (to load the data and then run the workload tests, respectively). We fix the data size to $1,000,000$ with Zipfian distribution of key popularity. We increase the number of clients from 1 to 100 to find out a saturation point of each targeted/scaled throughput for the settings. Here, we only present workload A (throughput vs average latency for the read and update); the other workloads display similar behavior.

As shown in Figure 10b, the client latencies of the DynamoRIO and RATEL settings for a given throughput are slightly similar until approximately $10,000$ ops/sec while it nearly keeps unchanged on Linux until more than $40,000$ ops/sec. Specifically, RATEL jitters until it achieves maximum throughput around $17,000$ ops/sec, while DynamoRIO is flat until $15,000$ ops/sec (the maximum is $21,000$ operations per second). The shared reason of the deceleration for both is that DynamoRIO slows down the speed of *Read* and *Update*. For RATEL, the additional bottleneck is the high frequency of lock contention with spin-lock primitive. For e.g., RATEL costs $18,320,000$ ms while DynamoRIO's the futex calls cost only around 500 ms for a given throughput of 10000 with 10 clients.